\documentclass[twocolumn]{aastex63}

\usepackage{xspace}
\usepackage[normalem]{ulem}

\usepackage{soul}

\usepackage{amsmath}
\usepackage{rotating}

\graphicspath{{./}{figures/}}

 \submitjournal{ApJL}
\shorttitle{HAWC Search for High-Mass Microquasars}
\shortauthors{HAWC Collaboration}

\begin{document}


\title{HAWC Search for High-Mass Microquasars}

\correspondingauthor{Chang Dong Rho}
\email{cdr397@uos.ac.kr}
\correspondingauthor{Ke Fang}
\email{kefang@physics.wisc.edu}

\author{A.~Albert}
\affiliation{Physics Division, Los Alamos National Laboratory, Los Alamos, NM, USA}

\author{R.~Alfaro}
\affiliation{Instituto de F\'{i}sica, Universidad Nacional Aut{\'o}noma de M{\'e}xico, Ciudad de Mexico, Mexico}

\author{C.~Alvarez}
\affiliation{Universidad Aut{\'o}noma de Chiapas, Tuxtla Guti{\'e}rrez, Chiapas, Mexico}

\author{J.R.~Angeles Camacho}
\affiliation{Universidad Aut{\'o}noma de Chiapas, Tuxtla Guti{\'e}rrez, Chiapas, Mexico}

\author{J.C.~Arteaga-Vel{\'a}zquez}
\affiliation{Universidad Michoacana de San Nicol{\'a}s de Hidalgo, Morelia, Mexico}

\author{K.P.~Arunbabu}
\affiliation{Instituto de Geof\'{i}sica, Universidad Nacional Aut{\'o}noma de M{\'e}xico, Ciudad de Mexico, Mexico}

\author{D.~Avila Rojas}
\affiliation{Instituto de F\'{i}sica, Universidad Nacional Aut{\'o}noma de M{\'e}xico, Ciudad de Mexico, Mexico}

\author{H.A.~Ayala Solares}
\affiliation{Department of Physics, Pennsylvania State University, University Park, PA, USA}

\author{V.~Baghmanyan}
\affiliation{Institute of Nuclear Physics Polish Academy of Sciences, PL-31342 IFJ-PAN, Krakow, Poland}

\author{E.~Belmont-Moreno}
\affiliation{Instituto de F\'{i}sica, Universidad Nacional Aut{\'o}noma de M{\'e}xico, Ciudad de Mexico, Mexico}

\author{S.Y.~BenZvi}
\affiliation{Department of Physics \& Astronomy, University of Rochester, Rochester, NY , USA}

\author{C.~Brisbois}
\affiliation{Department of Physics, University of Maryland, College Park, MD, USA}

\author{K.S.~Caballero-Mora}
\affiliation{Universidad Aut{\'o}noma de Chiapas, Tuxtla Guti{\'e}rrez, Chiapas, Mexico}

\author{T.~Capistr{\'a}n}
\affiliation{Instituto de Astronom\'{i}a, Universidad Nacional Aut{\'o}noma de M{\'e}xico, Ciudad de Mexico, Mexico}

\author{A.~Carrami{\~n}ana}
\affiliation{Instituto Nacional de Astrof\'{i}sica, {\'O}ptica y Electr{\'o}nica, Puebla, Mexico}

\author{S.~Casanova}
\affiliation{Institute of Nuclear Physics Polish Academy of Sciences, PL-31342 IFJ-PAN, Krakow, Poland}

\author{U.~Cotti}
\affiliation{Universidad Michoacana de San Nicol{\'a}s de Hidalgo, Morelia, Mexico}

\author{J.~Cotzomi}
\affiliation{Facultad de Ciencias F\'{i}sico Matem{\'a}ticas, Benem{\'e}rita Universidad Aut{\'o}noma de Puebla, Puebla, Mexico}

\author{E.~De la Fuente}
\affiliation{Departamento de F\'{i}sica, Centro Universitario de Ciencias Exactase Ingenierias, Universidad de Guadalajara, Guadalajara, Mexico}

\author{C.~de Le{\'o}n}
\affiliation{Universidad Michoacana de San Nicol{\'a}s de Hidalgo, Morelia, Mexico}

\author{R.~Diaz Hernandez}
\affiliation{Instituto Nacional de Astrof\'{i}sica, {\'O}ptica y Electr{\'o}nica, Puebla, Mexico}

\author{J.C.~D\'{i}az-V\'{e}lez}
\affiliation{Departamento de F\'{i}sica, Centro Universitario de Ciencias Exactase Ingenierias, Universidad de Guadalajara, Guadalajara, Mexico}

\author{B.L.~Dingus}
\affiliation{Physics Division, Los Alamos National Laboratory, Los Alamos, NM, USA}

\author{M.~Durocher}
\affiliation{Physics Division, Los Alamos National Laboratory, Los Alamos, NM, USA}

\author{M.A.~DuVernois}
\affiliation{Department of Physics, University of Wisconsin-Madison, Madison, WI, USA}

\author{R.W.~Ellsworth}
\affiliation{Department of Physics, University of Maryland, College Park, MD, USA}

\author{C.~Espinoza}
\affiliation{Instituto de F\'{i}sica, Universidad Nacional Aut{\'o}noma de M{\'e}xico, Ciudad de Mexico, Mexico}

\author{K.L.~Fan}
\affiliation{Department of Physics, University of Maryland, College Park, MD, USA}

\author{K.~Fang}
\affil{Kavli Institute for Particle Astrophysics and Cosmology (KIPAC), Stanford University, Stanford, CA 94305, USA}
\affil{NHFP Einstein Fellow}
\affil{Department of Physics, University of Wisconsin-Madison, Madison, WI, USA}

\author{N.~Fraija}
\affiliation{Instituto de Astronom\'{i}a, Universidad Nacional Aut{\'o}noma de M{\'e}xico, Ciudad de Mexico, Mexico}

\author{A.~Galv{\'a}n-G{\'a}mez}
\affiliation{Instituto de Astronom\'{i}a, Universidad Nacional Aut{\'o}noma de M{\'e}xico, Ciudad de Mexico, Mexico}

\author{J.A.~Garc{\'i}a-Gonz{\'a}lez}
\affiliation{Tecnologico de Monterrey, Escuela de Ingenier\'{i}a y Ciencias, Ave. Eugenio Garza Sada 2501, Monterrey, N.L., Mexico, 64849}

\author{F.~Garfias}
\affiliation{Instituto de Astronom\'{i}a, Universidad Nacional Aut{\'o}noma de M{\'e}xico, Ciudad de Mexico, Mexico}

\author{M.M.~Gonz{\'a}lez}
\affiliation{Instituto de Astronom\'{i}a, Universidad Nacional Aut{\'o}noma de M{\'e}xico, Ciudad de Mexico, Mexico}

\author{J.A.~Goodman}
\affiliation{Department of Physics, University of Maryland, College Park, MD, USA}

\author{J.P.~Harding}
\affiliation{Physics Division, Los Alamos National Laboratory, Los Alamos, NM, USA}

\author{S.~Hernandez}
\affiliation{Instituto de F\'{i}sica, Universidad Nacional Aut{\'o}noma de M{\'e}xico, Ciudad de Mexico, Mexico}

\author{B.~Hona}
\affiliation{Department of Physics and Astronomy, University of Utah, Salt Lake City, UT, USA}

\author{D.~Huang}
\affiliation{Department of Physics, Michigan Technological University, Houghton, MI, USA}

\author{F.~Hueyotl-Zahuantitla}
\affiliation{Universidad Aut{\'o}noma de Chiapas, Tuxtla Guti{\'e}rrez, Chiapas, Mexico}

\author{P.~H{\"u}ntemeyer}
\affiliation{Department of Physics, Michigan Technological University, Houghton, MI, USA}

\author{A.~Iriarte}
\affiliation{Instituto de Astronom\'{i}a, Universidad Nacional Aut{\'o}noma de M{\'e}xico, Ciudad de Mexico, Mexico}

\author{A.~Jardin-Blicq}
\affiliation{Max-Planck Institute for Nuclear Physics, 69117 Heidelberg, Germany}
\affiliation{Department of Physics, Faculty of Science, Chulalongkorn University, 254 Phayathai Road, Pathumwan, Bangkok 10330, Thailand}
\affiliation{National Astronomical Research Institute of Thailand (Public Organization), Don Kaeo, MaeRim, Chiang Mai 50180, Thailand}

\author{V.~Joshi}
\affiliation{Erlangen Centre for Astroparticle Physics, Friedrich-Alexander-Universit{\"a}t Erlangen-N{\"u}rnberg, Erlangen, Germany}

\author{D.~Kieda}
\affiliation{Department of Physics and Astronomy, University of Utah, Salt Lake City, UT, USA}

\author{A.~Lara}
\affiliation{Instituto de Geof\'{i}sica, Universidad Nacional Aut{\'o}noma de M{\'e}xico, Ciudad de Mexico, Mexico}

\author{J.~Lee}
\affiliation{Natural Science Research Institute, University of Seoul, Seoul, Republic of Korea}

\author{W.H.~Lee}
\affiliation{Instituto de Astronom\'{i}a, Universidad Nacional Aut{\'o}noma de M{\'e}xico, Ciudad de Mexico, Mexico}

\author{H.~Le{\'o}n Vargas}
\affiliation{Instituto de F\'{i}sica, Universidad Nacional Aut{\'o}noma de M{\'e}xico, Ciudad de Mexico, Mexico}

\author{J.T.~Linnemann}
\affiliation{Department of Physics and Astronomy, Michigan State University, East Lansing, MI, USA}

\author{A.L.~Longinotti}
\affiliation{Instituto de Astronom\'{i}a, Universidad Nacional Aut{\'o}noma de M{\'e}xico, Ciudad de Mexico, Mexico}

\author{G.~Luis-Raya}
\affiliation{Universidad Politecnica de Pachuca, Pachuca, Hgo, Mexico}

\author{J.~Lundeen}
\affiliation{Department of Physics and Astronomy, Michigan State University, East Lansing, MI, USA}

\author{K.~Malone}
\affiliation{Physics Division, Los Alamos National Laboratory, Los Alamos, NM, USA}

\author{O.~Martinez}
\affiliation{Facultad de Ciencias F\'{i}sico Matem{\'a}ticas, Benem{\'e}rita Universidad Aut{\'o}noma de Puebla, Puebla, Mexico}

\author{J.~Mart{\'i}nez-Castro}
\affiliation{Centro de Investigaci\'on en Computaci\'on, Instituto Polit\'ecnico Nacional, Ciudad de Mexico, Mexico}

\author{J.A.~Matthews}
\affiliation{Dept of Physics and Astronomy, University of New Mexico, Albuquerque, NM, USA}

\author{P.~Miranda-Romagnoli}
\affiliation{Universidad Aut{\'o}noma del Estado de Hidalgo, Pachuca, Mexico}

\author{J.A.~Morales-Soto}
\affiliation{Universidad Michoacana de San Nicol{\'a}s de Hidalgo, Morelia, Mexico}

\author{E.~Moreno}
\affiliation{Facultad de Ciencias F\'{i}sico Matem{\'a}ticas, Benem{\'e}rita Universidad Aut{\'o}noma de Puebla, Puebla, Mexico}

\author{M.~Mostaf{\'a}}
\affiliation{Department of Physics, Pennsylvania State University, University Park, PA, USA}

\author{A.~Nayerhoda}
\affiliation{Institute of Nuclear Physics Polish Academy of Sciences, PL-31342 IFJ-PAN, Krakow, Poland}

\author{L.~Nellen}
\affiliation{Instituto de Ciencias Nucleares, Universidad Nacional Aut{\'o}noma de Mexico, Mexico City, Mexico}

\author{M.~Newbold}
\affiliation{Department of Physics and Astronomy, University of Utah, Salt Lake City, UT, USA}

\author{M.U.~Nisa}
\affiliation{Department of Physics and Astronomy, Michigan State University, East Lansing, MI, USA}

\author{R.~Noriega-Papaqui}
\affiliation{Universidad Aut{\'o}noma del Estado de Hidalgo, Pachuca, Mexico}

\author{L.~Olivera-Nieto}
\affiliation{Max-Planck Institute for Nuclear Physics, 69117 Heidelberg, Germany}

\author{N.~Omodei}
\affiliation{Department of Physics, Stanford University: Stanford, CA 94305–4060, USA}

\author{A.~Peisker}
\affiliation{Department of Physics and Astronomy, Michigan State University, East Lansing, MI, USA}

\author{Y.~P{\'e}rez Araujo}
\affiliation{Instituto de Astronom\'{i}a, Universidad Nacional Aut{\'o}noma de M{\'e}xico, Ciudad de Mexico, Mexico}

\author{C.D.~Rho}
\affiliation{Natural Science Research Institute, University of Seoul, Seoul, Republic of Korea}

\author{Y.J.~Roh}
\affiliation{Natural Science Research Institute, University of Seoul, Seoul, Republic of Korea}

\author{D.~Rosa-Gonz{\'a}lez}
\affiliation{Instituto Nacional de Astrof\'{i}sica, {\'O}ptica y Electr{\'o}nica, Puebla, Mexico}

\author{F.~Salesa Greus}
\affiliation{Institute of Nuclear Physics Polish Academy of Sciences, PL-31342 IFJ-PAN, Krakow, Poland}
\affiliation{Instituto de F\'{i}sica Corpuscular, CSIC, Universitat de Val\`{e}ncia, E-46980, Paterna, Valencia, Spain}

\author{A.~Sandoval}
\affiliation{Instituto de F\'{i}sica, Universidad Nacional Aut{\'o}noma de M{\'e}xico, Ciudad de Mexico, Mexico}

\author{M.~Schneider}
\affiliation{Department of Physics, University of Maryland, College Park, MD, USA}

\author{J.~Serna-Franco}
\affiliation{Instituto de F\'{i}sica, Universidad Nacional Aut{\'o}noma de M{\'e}xico, Ciudad de Mexico, Mexico}

\author{A.J.~Smith}
\affiliation{Department of Physics, University of Maryland, College Park, MD, USA}

\author{R.W.~Springer}
\affiliation{Department of Physics and Astronomy, University of Utah, Salt Lake City, UT, USA}

\author{K.~Tollefson}
\affiliation{Department of Physics and Astronomy, Michigan State University, East Lansing, MI, USA}

\author{I.~Torres}
\affiliation{Instituto Nacional de Astrof\'{i}sica, {\'O}ptica y Electr{\'o}nica, Puebla, Mexico}

\author{R.~Torres-Escobedo}
\affiliation{Departamento de F\'{i}sica, Centro Universitario de Ciencias Exactase Ingenierias, Universidad de Guadalajara, Guadalajara, Mexico}

\author{R.~Turner}
\affiliation{Department of Physics, Michigan Technological University, Houghton, MI, USA}

\author{F.~Ure{\~n}a-Mena}
\affiliation{Instituto Nacional de Astrof\'{i}sica, {\'O}ptica y Electr{\'o}nica, Puebla, Mexico}

\author{L.~Villase{\~n}or}
\affiliation{Facultad de Ciencias F\'{i}sico Matem{\'a}ticas, Benem{\'e}rita Universidad Aut{\'o}noma de Puebla, Puebla, Mexico }

\author{I.J.~Watson}
\affiliation{Natural Science Research Institute, University of Seoul, Seoul, Republic of Korea}

\author{T.~Weisgarber}
\affiliation{Department of Physics, University of Wisconsin-Madison, Madison, WI, USA}

\author{E.~Willox}
\affiliation{Department of Physics, University of Maryland, College Park, MD, USA}

\author{H.~Zhou}
\affiliation{Tsung-Dao Lee Institute \& School of Physics and Astronomy, Shanghai Jiao Tong University, Shanghai, China}

\begin{abstract}
Microquasars with high-mass companion stars are promising very-high-energy (VHE; 0.1-100~TeV) gamma-ray emitters, but their behaviors above 10~TeV are poorly known. Using the High Altitude Water Cherenkov (HAWC) observatory, we search for excess gamma-ray emission coincident with the positions of known high-mass microquasars (HMMQs). No significant emission is observed for LS~5039, Cygnus X-1, Cygnus X-3, and SS~433 with 1,523 days of HAWC data. We set the most stringent limit above 10~TeV obtained to date on each individual source. Under the assumption that HMMQs produce gamma rays via a common mechanism, we have  performed source-stacking searches, considering two different scenarios: I) gamma-ray luminosity is a fraction $\epsilon_\gamma$ of the microquasar jet luminosity, and II) very-high-energy gamma rays are produced by relativistic electrons up-scattering the radiation field of the companion star in a magnetic field $B$. We obtain $\epsilon_\gamma < 5.4\times 10^{-6}$ for scenario I, which tightly constrains models that suggest observable high-energy neutrino emission by HMMQs. In the case of scenario II, the non-detection of VHE gamma rays yields a strong magnetic field, which challenges synchrotron radiation as the dominant mechanism of the microquasar emission between 10~keV and 10~MeV. 
\end{abstract}

\keywords{
Gamma-ray sources(633), High mass x-ray binary stars (733)
}


\section{Introduction}

Microquasars are radio emitting X-ray binaries (XRBs) with relativistic outflows or jets \citep{1999ARA&A..37..409M}. Powered by stellar-mass compact objects, they mimic extragalactic quasars on smaller scales and present accretion and formation of jets. 
Microquasars with high-mass companion stars (or, high-mass microquasars, HMMQs) share many similarities in geometry and observational behaviors  \citep{2002A&A...394..193P}.
A typical HMMQ has a young O or B type star with mass greater than $10\,M_\odot$, and experiences mass transfer between the companion and the compact object via stellar winds. In addition, they usually show persistent radio emission. 

HMMQs are suggested to be promising TeV $\gamma$-ray emitters (\citealt{Marcote:2015abc}, also see the review by \citealt{2013A&ARv..21...64D} and the references therein). Indeed, a few of them have been observed in high-energy (HE; 0.1-100~GeV) and/or very-high-energy (0.1-100~TeV) gamma rays, including  LS~I~+61$^{\circ}$~303 \citep{2006Sci...312.1771A}, LS~5039 \citep{Mariaud:2015pha}, Cygnus~X-3 (only in HE; \citealp{10.2307/27736637}), and Cygnus~X-1(possibly only in HE; \citealp{2017MNRAS.471.3657Z}). Although not all HMMQs are detected in gamma rays, the HMMQ branch of the gamma-ray binaries raises an interesting question as to whether $\gamma$-ray emission is a common feature in the HMMQ population. 

The $\gamma$-ray production mechanism of the known binary systems is largely unknown. The emission has been suggested to be produced by either the accretion-powered microquasar jets and outflows or the rotation-powered pulsar winds \citep{2013A&ARv..21...64D}. The origin of the gamma rays is also debated to be either from the decay of neutral pions via hadronic interactions or from the inverse Compton scattering of optical-to-UV photons from the donor star by relativistic electrons. 

Motivated by these questions, we search for VHE $\gamma$-ray emission from HMMQs using the High Altitude Water Cherenkov (HAWC) observatory.
 HAWC observes VHE gamma rays via the induced extensive air showers produced from a series of pair production and Bremsstrahlung. It provides an unprecedented sensitivity for the observation of VHE gamma rays above $\sim 10$~TeV. 
For each source in our target list, we derive upper limits on the VHE emission and compare these to existing multi-wavelength observational data of the source.  
By stacking the likelihoods of the fitted $\gamma$-ray emission from all known HMMQs accessible to HAWC, assuming that they produce gamma rays via a common mechanism, 
the absence of detection strongly constrains the VHE emission efficiency and the magnetic field strength in the relativistic outflows of microquasars.

This work is different from \citet{Abeysekara:2018qtj} where VHE $\gamma$-ray emission from the extended jets of SS~433 is studied. Here, we focus on the gamma-ray emission in the vicinity of the binary system with a size on the order of $\sim 0.1$ astronomical unit (AU).

This paper is organized as follows. In $\S\ref{sec:methods}$ we describe the methods of our analysis including the construction of the target source list $\S\ref{subsec:sourcelist}$, the analysis of HAWC data $\S\ref{subsec:hawcdata}$, and the stacking of likelihoods $\S\ref{subsec:stacking}$. The results are presented in $\S\ref{sec:results}$ and discussed in $\S\ref{sec:discussion}$.

\section{Methods}\label{sec:methods}
\subsection{Source Selection}\label{subsec:sourcelist}

\begin{deluxetable*}{cccccccc}
\tablecaption{List of high-mass microquasars in the HAWC FOV and their properties, including the location (RA, DEC) and distance $D$ of the binary system, the companion star's temperature $T_*$, radius $R_*$, separation from the compact object $d_*$, and the compact object's jet power $L_{\rm jet}$.   \label{tab:RHMXB}}
\tablewidth{700pt}
\tabletypesize{\scriptsize}
\tablehead{
\colhead{Name} & \colhead{RA} & \colhead{DEC}  & \colhead{$T_*$} & \colhead{$R_*$} & \colhead{$d_*$} & \colhead{Jet kinetic power $L_{\rm jet}$ } & \colhead{Distance $D$}  \\ 
\colhead{} & \colhead{} & \colhead{} &  \colhead{[$10^4$~K]} & \colhead{[$R_\odot$]} & \colhead{[AU]} &  \colhead{[$\rm erg\,s^{-1}$]} & \colhead{[kpc]}
} 
\startdata
   LS 5039\tablenotemark{\scriptsize a}  \space     & 18:26:15.1      & -14$^{\circ}$50'54''      & 3.9 &  9.3 & 0.1 & $10^{36}$~\tablenotemark{\scriptsize e} & 2.9 \\
    CYG X-1\tablenotemark{\scriptsize b} \space     & 19:58:21.7      & +35$^{\circ}$12'06''      & 3.1 &  20  & 0.2 &  $(4-14)\times10^{36}$ &  2.2 \\
    CYG X-3\tablenotemark{\scriptsize c} \space     & 20:32:26.5      & +40$^{\circ}$57'09''      & 4-5 &  $<$2 & 0.02 & $10^{38}$ & 7.0 \\
    SS 433\tablenotemark{\scriptsize d} \space      & 19:11:49.6      & +04$^{\circ}$58'58''      & 3.25 &  5.5\tablenotemark{\scriptsize f} & 0.5  & $10^{39}$ & 5.5 \\
\enddata

\tablenotetext{a}{\citealt{Casares:2005ig, 2006AA...451..259P}}
\tablenotetext{b}{\citealt{2005Natur.436..819G, 2006ApJ...636..316H, 2007MNRAS.376.1341R, 2014MNRAS.440L..61Z}}
\tablenotetext{c}{\citealt{2012MNRAS.421.2956Z, Koljonen:2017gah}}
\tablenotetext{d}{\citealt{1986ApJ...308..152W, Begelman:2006bi}}
\tablenotetext{e}{It has also been suggested that $\gamma$-ray emission in this source is powered by pulsar winds \citep{Dubus:2006ze}.}
\tablenotetext{f}{Based on the mass of the donor star $12.3\,M_\odot$ \citep{Kubota_2010} and the  stellar mass-radius relation \citep{Eker_2018}.}

\end{deluxetable*}

We select target sources based on two criteria: i) it is a confirmed X-ray binary system with steady radio emission, i.e. a microquasar, within the sky coverage of HAWC, and ii) it does not present transient X-ray outbursts like XTE~J0421+560 \citep{Frontera:1998jp}. 
Applying these conditions to the high-mass X-ray binary catalog \citep{Liu:2007tu},  we are left with four HMMQs as target sources: LS~5039, Cygnux~X-1, Cygnus~X-3, and SS~433. Although LS~I~+61$^{\circ}$~303 may also seem to satisfy our criteria, it is at the edge of the HAWC field-of-view (FOV). Due to the poor detector sensitivity in that region, we do not include LS~I~+61$^{\circ}$~303 in the target list. 

Table~\ref{tab:RHMXB} lists the relevant properties of the four HMMQs studied in this paper.

\subsection{HAWC Analysis}\label{subsec:hawcdata}
HAWC is a high duty cycle, wide field-of-view particle sampling array consisting of 300 water Cherenkov detectors (WCDs) covering a combined geometrical area of $\sim22,000$ m$^2$ \citep{Abeysekara:2017mjj}.
It is located at a latitude of $\sim19^{\circ}$ N and at an altitude of $\sim4,100$ meters in Mexico. Each WCD contains 200,000 litres of purified water and four upward-facing photomultiplier tubes are anchored to the bottom \citep{Abeysekara:2017mjj}.
The dataset used in this analysis consists of cumulative observational data averaged over the time period of 1,523 days and the energy of the $\gamma$-ray events is estimated from the number of hit PMTs per gamma-ray event. The expected energy and angular resolutions are $\geq20\%$ and $\geq0.1^{\circ}$, respectively, based on the Crab Nebula analysis \citep{Abeysekara:2017mjj} where more details about the HAWC setup, data, and general source analysis procedures can also be found.

\begin{deluxetable}{ccc}
\tablecaption{Quasi-differential energy bins.\label{tab:quasi}}
\tablewidth{700pt}
\tabletypesize{\scriptsize}
\tablehead{
\colhead{Energy Bin} & \colhead{Energy Range} & \colhead{Pivot Energy} \\
& \colhead{[TeV]} & \colhead{[TeV]}
} 
\startdata
        1 & 1.0-3.2 & 1.8 \\
        2 & 3.2-10.0 & 5.6 \\
        3 & 10.0-31.6 & 17.8 \\
        4 & $>$31.6 & 56.2 \\
\enddata
\end{deluxetable}

Likelihood fitting with given spatial and spectral models is used to compute the $\gamma$-ray energy spectrum. 
In each energy bin $i$, a simple power-law spectral model is used to describe the $\gamma$-ray spectrum,

\begin{equation}
    \Phi_i=A_i\left(\frac{E}{E_{i, \text{piv}}}\right)^{-\alpha_i},
\end{equation}
where $\Phi_i$ is the differential flux at the pivot energy $E_{i, \text{piv}}$, $A_i$ is the flux normalization, $E$ is the photon energy, and $\alpha_i$ is the spectral index. 
For this analysis, we use four quasi-differential energy bins as listed in Table~\ref{tab:quasi}. Within each bin, we adopt a spectral index $\alpha_i = 2.7$, which is a good approximation for point-like HAWC sources \citep{Abeysekara:2017hyn}. The systematic uncertainties due to the unknown spectral index and detector response functions will be discussed below.
Since the binary systems have a typical size of $0.1$~AU and are located at a distance of several kilo-parsecs, a point-source morphology is adopted for all target sources.

\begin{figure*}[htp]
    \centering
    \includegraphics[width=.49\textwidth]{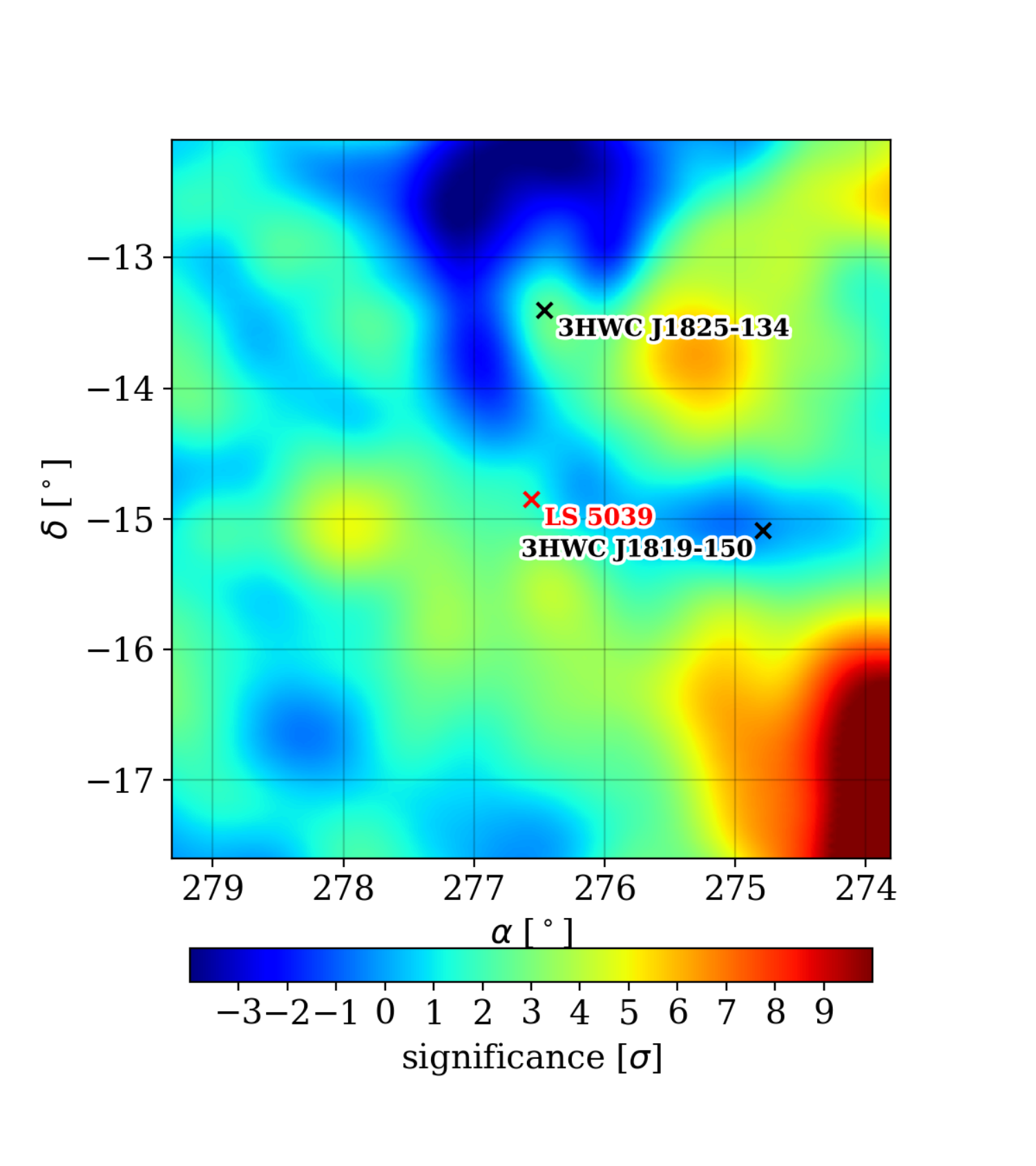}
    \hfill
    \includegraphics[width=.49\textwidth]{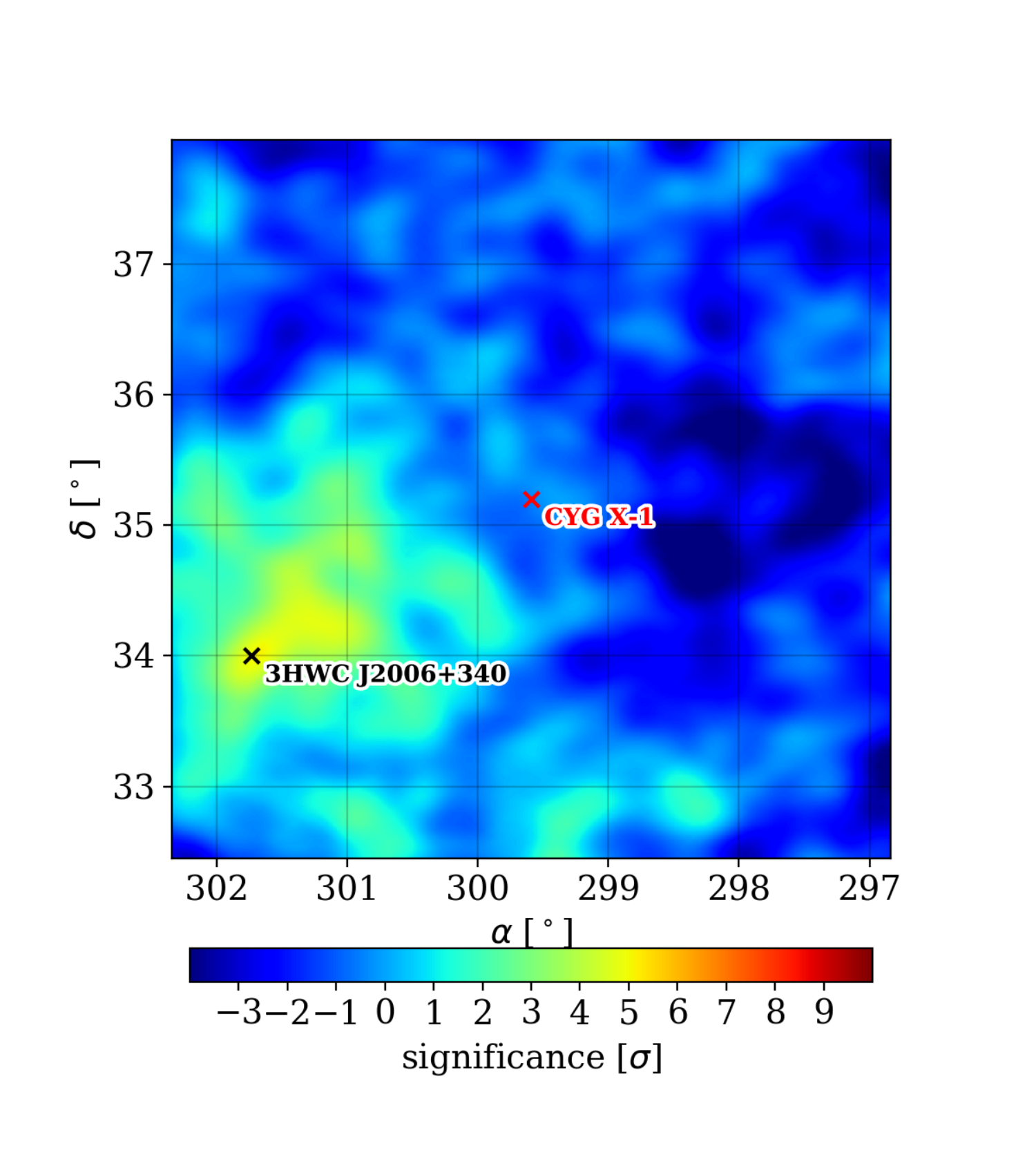}\\
    \includegraphics[width=.49\textwidth]{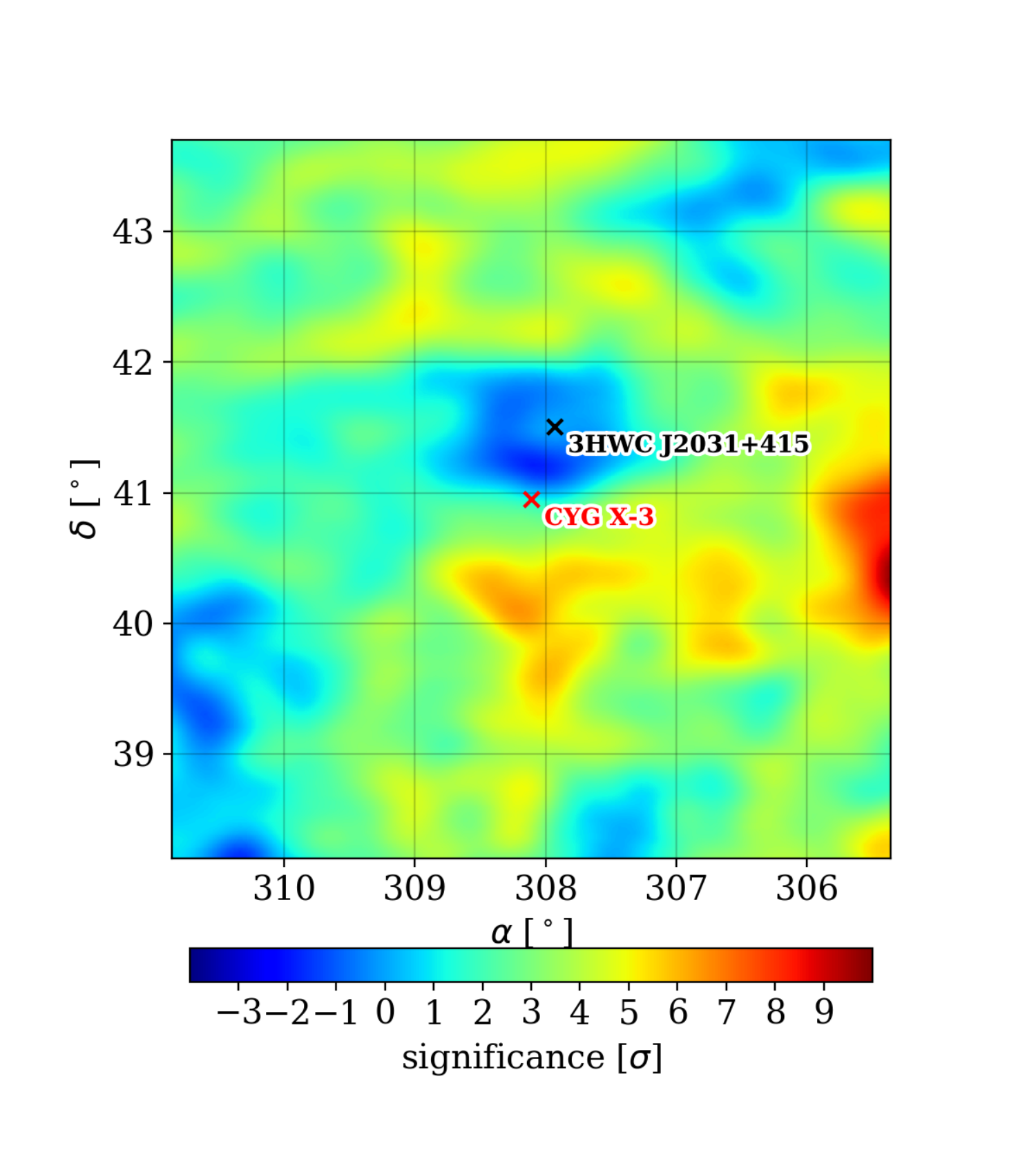}
    \hfill
    \includegraphics[width=.49\textwidth]{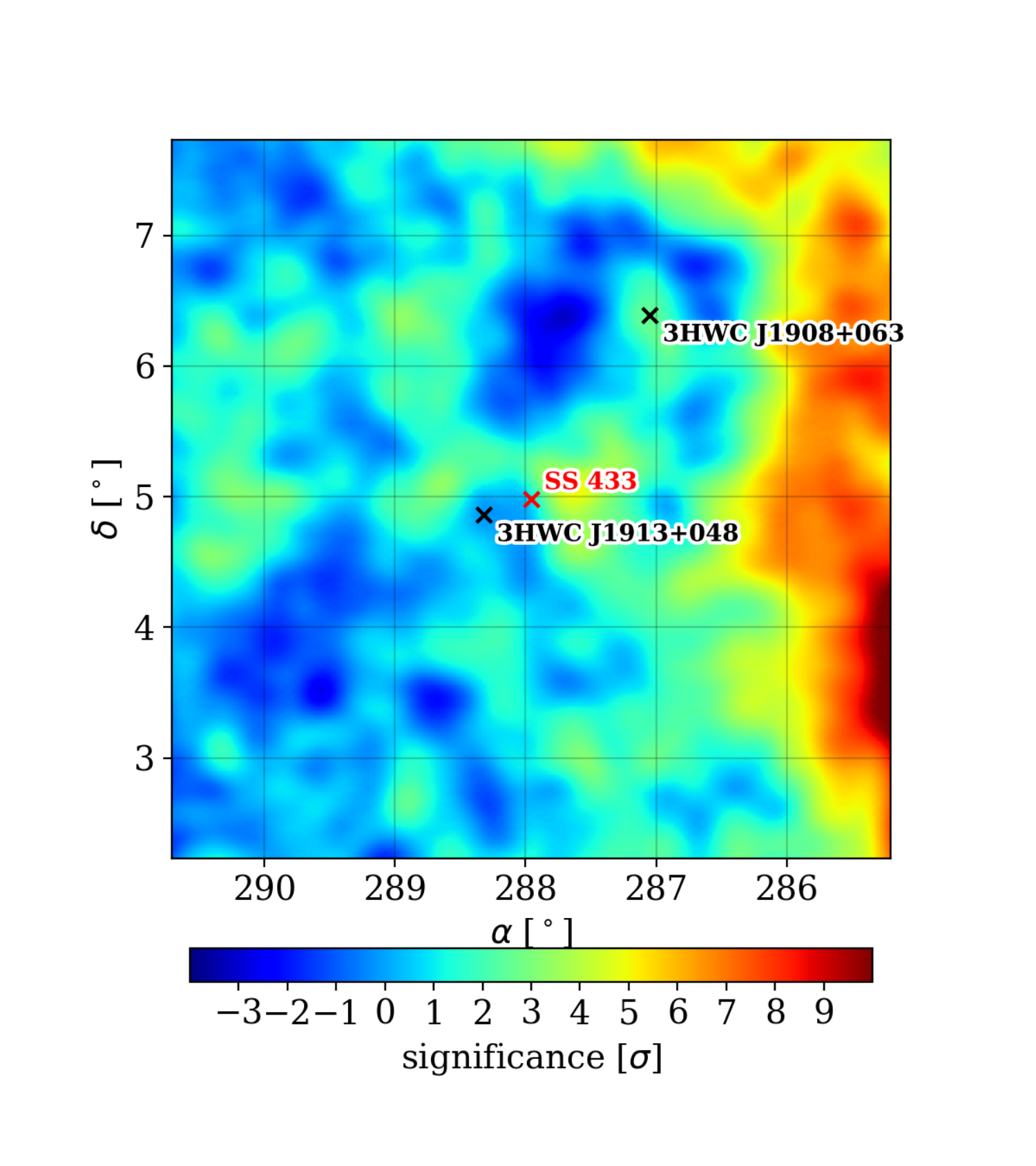}\\
    \caption{\label{fig:resmap} Residual significance maps of the regions centered around LS 5039 (top left), CYG X-1 (top right), CYG X-3 (bottom left), and SS 433 (bottom right) produced using 1,523 days of HAWC data. We also show in these maps the labelled 3HWC sources fitted and subtracted. These significance maps have been made by fitting, per pixel, an $E^{-2.7}$ spectrum and a point-like source morphology.}
\end{figure*}

All four target sources are located in source-confused regions close to the Galactic plane with several nearby TeV sources. The HAWC significance maps of each region, with nearby sources labeled, are shown in Appendix $\ref{subsec:sigmap}$, Figure~\ref{fig:sigmap}. The residual maps, as shown in Figure~\ref{fig:resmap}, are obtained with the following steps.
We first fit background sources using their known locations and spectral indices from the 3HWC Catalog \citep{2020arXiv200708582A}. 
In particular, we fit point-like background sources such as 3HWC~J1819-150 and 3HWC~J1913+048 with a point source model. The regions of interest also contain four extended sources. We use a  simple Gaussian morphology for 3HWC~J2006+340 and 3HWC~J1908+063. The 3HWC~J1825-134 area consists of two pulsar wind nebulae, HESS~J1825-137 and HESS~J1826-130 \citep{Abdalla:2018qgt}, positioned above and below the location of 3HWC~J1825-134. Hence, we apply an asymmetrical Gaussian morphological model to 3HWC~J1825-134 with its semi-major axis positioned along the line joining the three VHE source locations. 
Finally, the Cygnus cocoon's gamma-ray profile is ``flat'' \citep{Hona:2019ysf}. Hence we adopt a disk-like morphological model for 3HWC~J2031+415.

The obtained best-fit models for the nearby sources are then subtracted from the original HAWC 1,523 transit maps to produce the residual maps as shown in Figure~\ref{fig:resmap}. Then, we fit for the flux normalization of each HMMQ to find their flux upper limits.

We calculate a test statistic (TS) for $\gamma$-ray detection based on the logarithm of the likelihood ratio when fitting with the residual maps with and without the target source in all energy bins,

\begin{equation}
   {\rm TS} \equiv 2\,\left[\ln {\mathcal{L}}(\hat{A}) - \ln  {\mathcal{L}}(A=0)\right],
    \label{eqn:TS}
\end{equation}
where $\mathcal{L}$ is the Poisson likelihood function and $\hat{A}$ is the best-fit normalization found from the maximum-likelihood estimators.
We obtain \textit{a priori} statistical significance for a given location in the sky via,

\begin{equation}
    \sigma \approx \pm\sqrt{\rm{TS}}. 
    \label{eqn:signif}
\end{equation}

The best-fit normalization,
$\hat{A}$, is used as an input to a Markov-Chain
Monte Carlo (MCMC). MCMC then estimates the distribution of the posterior
likelihood around the maximum value of the likelihood with a positive uniform prior assumed. From the obtained MCMC distribution, we can finally compute the 95\% credible upper limit on the flux normalization.

The HAWC data analysis involves forward-folding of the assumed morphological and spectral models through the detector response to obtain the expected gamma-ray counts. Following \citet{Abeysekara:2019edl}, we evaluate the detector systematic uncertainties by applying various versions of the detector response. Also, different spectral indices between 2.0 and 3.0 with an interval of 0.1 are applied to study the source spectrum.
The systematic errors on the flux normalizations due to different detector responses and astrophysical spectral indices are computed for each source at each quasi-differential energy bin and for one full energy bin containing data from all four bins. The errors are shown in Table~\ref{tab:sys-quasi} in Appendix \ref{sec:systematics}.

\subsection{Stacking of Likelihoods}\label{subsec:stacking}
Due to their similarity in the source structure, such as the accretion disk--jet configuration, and in the radiation background, such as thermal photons from  donor stars of similar star type, temperature, and size, the HMMQ population could, in principle, produce gamma rays with one same mechanism \citep{2013A&ARv..21...64D}. By combining the observations of all HMMQs in the HAWC FOV, we can constrain the common factors that impact the $\gamma$-ray production in these microquasars.

Below we consider two generic models, referred to as scenarios I and II, for VHE $\gamma$-ray emission in microquasar jets. In the first scenario, we assume that $\gamma$-ray luminosity is proportional to the kinetic power of the jets,

\begin{equation}
    L_\gamma = \epsilon_\gamma\,L_{\rm jet}. 
\end{equation}
This is a general assumption which may be satisfied by different $\gamma$-ray production models such as neutral pion decay from hadronic interactions.  
The $\gamma$-ray flux in scenario I can be written as 

\begin{equation}\label{eqn:phi_gamma1}
    \Phi_\gamma = \frac{\epsilon_\gamma\,L_{\rm jet}}{4\pi\,D^2}\,K_p\,\left(\frac{E}{E_{\rm piv}}\right)^{-p},
\end{equation}
where $D$ is the distance to the source, $K_p = ({2-p})\,E_{\rm piv}^{-p}/({E_{\rm max}^{2-p} - E_{\rm min}^{2-p}})$ is a normalization factor for spectral index $p$, and $K_p = E_{\rm piv}^{-2}/\log(E_{\rm max}/E_{\rm min})$ for $p=2$. Also, $E_{\rm min} = 1$~TeV, and $E_{\rm max} = 100$~TeV are the boundaries of the energy bin used for the stacking analysis (instead of the quasi-differential bins used in Sec.~\ref{subsec:hawcdata}) with $E_{\rm piv}=7$~TeV as the pivot energy. $L_{\rm jet}$ and $D$ of the target sources are listed in Table~\ref{tab:RHMXB}. 

In the second scenario, we consider the model summarized in \citet{2013A&ARv..21...64D}, where gamma rays are produced when relativistic electrons accelerated by the jets upscatter optical photons from the donor star. See Appendix \ref{subsec:model} for more details regarding the modeling of $\gamma$-ray production. In this model, the inverse Compton emission of a HMMQ is expected to peak at TeV energies and the corresponding synchrotron emission is typically at $10$~keV--10~MeV. The energy flux of the two components, $F_{\rm syn}$ and $F_{\rm IC}$, are connected by

\begin{equation}\label{eqn:ratio}
     \frac{F_{\rm syn}}{F_{\rm IC}} \approx \frac{u_B}{u_0\,f_{\rm KN}}, 
\end{equation}
where $u_0$ is the energy density of the radiation field of the star (equation~\ref{eqn:u_0}) and $u_B = B^2/8\pi$ is the magnetic energy density. Since the stellar radiation field is in the optical band, the inverse Compton emission of VHE electrons are in the Klein-Nishina regime. The unitless $f_{\rm KN}$ factor, evaluated at the inverse Compton break energy, $E_{\rm IC, bk}$, (equation~\ref{eqn:FKN}) accounts for the suppression of the inverse Compton cross section.

The $\gamma$-ray flux in scenario II can be expressed as

\begin{equation}\label{eqn:phi_gamma2}
    \Phi_\gamma = \frac{F_{\rm syn}\,u_0\,f_{\rm KN}}{u_B}\,K_p\,\left(\frac{E}{E_{\rm piv}}\right)^{-p}.
\end{equation}
We estimate the synchrotron flux $F_{\rm syn}$ using the measured X-ray (or MeV $\gamma$-ray) energy flux, $F_{\rm obs,bk}$,  between $0.1 \, E_{\rm syn, bk}$ and $10\,E_{\rm syn, bk}$, where $E_{\rm syn, bk}$ (equation~\ref{eqn:E_synbk}) is the peak energy of the synchrotron emission suggested by models fitted to the multiwavelength data. The energy density of the radiation field $u_0$ is derived from observed properties, including stellar temperature, radius, and separation from the compact object as listed in Table~\ref{tab:RHMXB}. The $u_0$ and $f_{\rm KN}$ used in the analysis are listed in Table~\ref{tab:RHMXB_Derived} in Section~\ref{subsec:model} of the Appendix. 

In both scenarios, the $\gamma$-ray flux of the $i^{\text{th}}$ source can be written as

\begin{equation}
    \Phi_i = K \, C_i\, \left(\frac{E}{E_{\rm piv}}\right)^{-p}, 
\end{equation}
where $C_i$ is the source-dependent contribution factor\footnote{The contribution factor is sometimes referred to as the ``J-factor'' in the literature.}, $C_i = K_p\, L_{\rm jet, i} / 4\pi\,D_i^2$ in scenario I and $C_{i}= K_p\, u_{0,i}\,f_{\rm KN, i}\,F_{\rm syn, i}$ in scenario II. $K$ is the ``weighting factor'' shared between the sources, specifically, $K = \epsilon_\gamma$ in scenario I and $K = 1 / u_B$ in scenario II.

We perform a likelihood fit for each target HMMQ to obtain their best-fit flux normalization. The sources are then stacked in the likelihood space weighted by their relevant contribution factors $C_i$, 

\begin{equation}
    \ln {\cal L} (p, K) = \sum_i \ln {\cal L}_i (p, K, C_i).
\end{equation}
The credible interval of the ``linked'' flux normalization for a given $p$ is obtained  using  MCMC following the steps described in $\S\ref{subsec:hawcdata}$. 
By scanning the index $p$ between $2.0$ and $3.0$ with $0.1$ interval, we can find the upper limit of $\hat{K}$ and the best-fit $\hat{p}$ that corresponds to the peak of the maximum log-likelihood $\ln {\cal L}(p, \hat{K}(p))$. The largest difference in $K$ obtained when varying $p$ is used as an estimation of the statistical error due to the scanning of the index. The detector systematic uncertainties are evaluated by applying various versions of the detector response for the cases with the best-fit spectral index.

Finally, the stacked flux is used to derive the limits on the weighting factor $K$. Note that the stacked flux depends on the definition of the contribution factor in a physical model. The fluxes in different scenarios are not directly comparable.

\section{Results}\label{sec:results}
\subsection{Upper Limits on Individual Sources}

\begin{figure*}[htp]
    \centering
    \includegraphics[width=.49\textwidth]{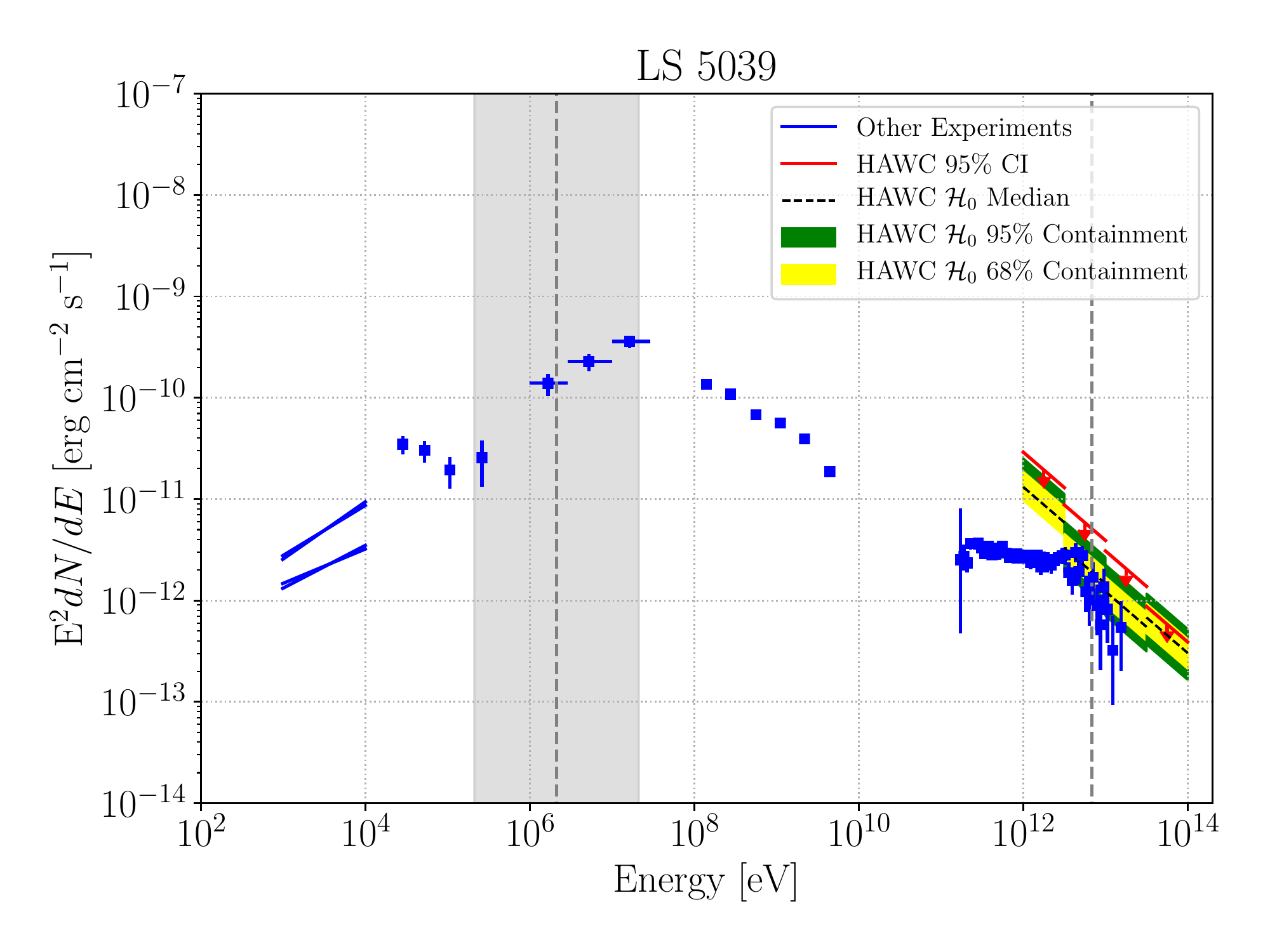}
    \hfill
    \includegraphics[width=.49\textwidth]{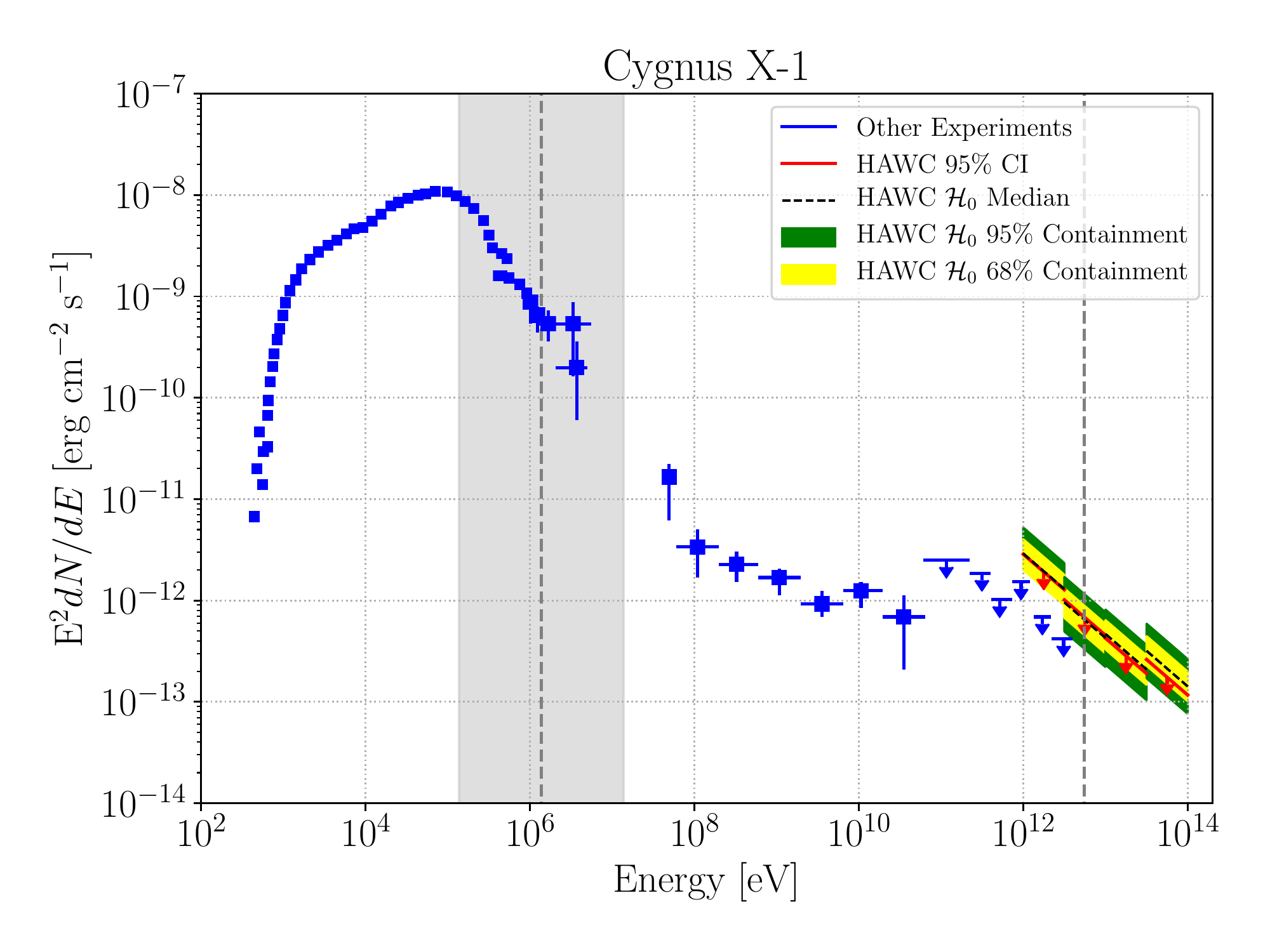}\\
    \includegraphics[width=.49\textwidth]{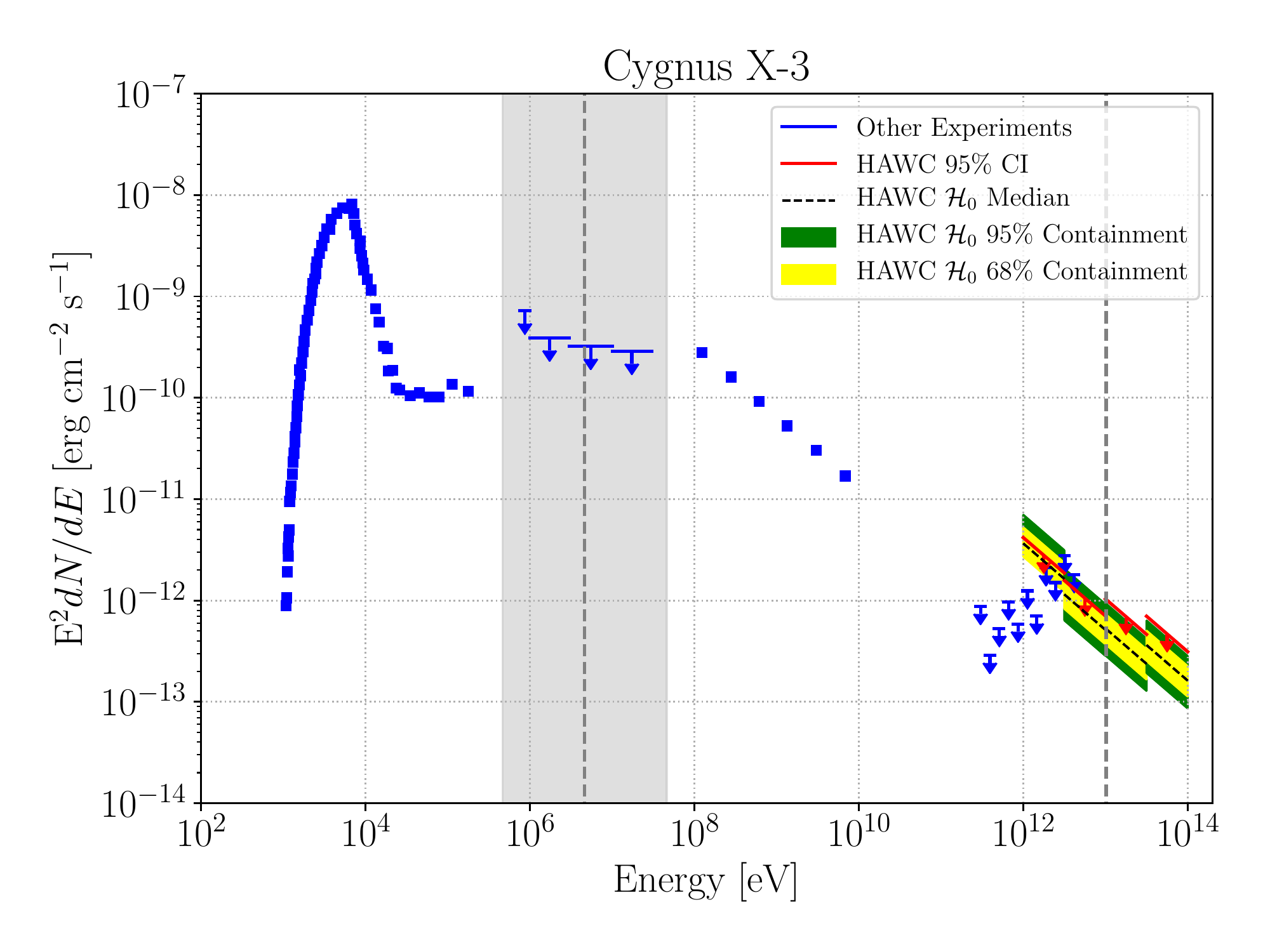}
    \hfill
    \includegraphics[width=.49\textwidth]{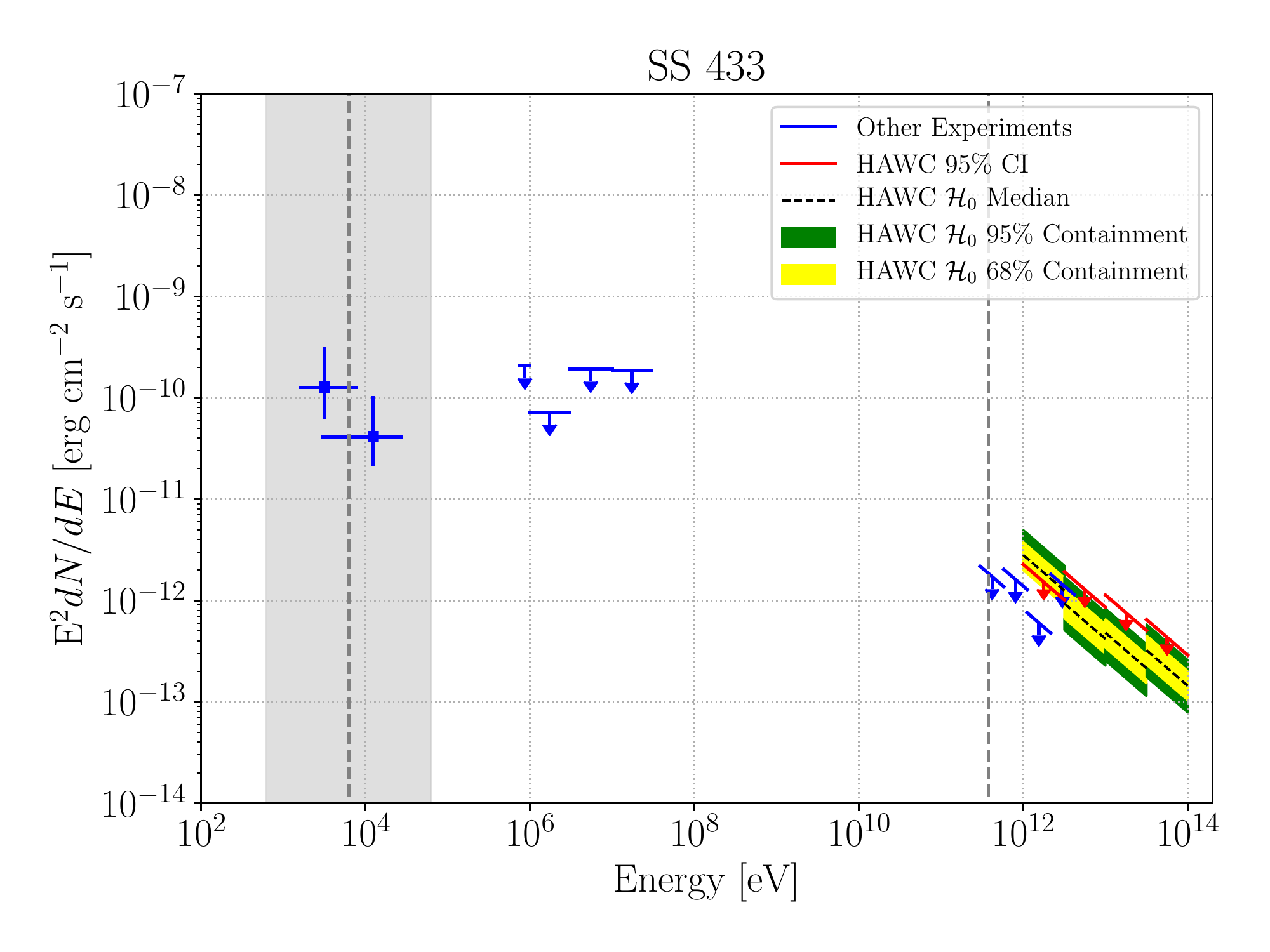}\\
    \caption{\label{fig:sed} Spectral energy distribution of LS~5039 (top left), CYG~X-1 (top right), CYG~X-3 (bottom left), and SS~433 (bottom right), in comparison with the upper limits on VHE $\gamma$-rays derived in this work from the HAWC observation. The blue data points below $\sim 0.1$~GeV correspond to the multi-wavelength data retrieved from other experiments, including \citet{Goldoni:2006rd, Jourdain_2011, Vilhu:2003rv, Cherepashchuk:2003tw, 2000A&AS..143..145S}. The high GeV to low TeV blue data points are the gamma-ray observation by various IACTs: LS~5039 by H.E.S.S. \citep{Mariaud:2015pha}, Cyg~X-1 by MAGIC \citep{2017MNRAS.471.3657Z}, Cyg~X-3 by VERITAS \citep{Archambault:2013yva}, and SS~433 by MAGIC and H.E.S.S. combined \citep{Ahnen:2017tsc}. The red upper limits are the 95\% HAWC quasi-differential credible intervals for each HMMQ.
    The vertical grey dashed lines correspond to characteristic synchrotron and inverse Compton energies, $E_{\rm syn, bk}$ and  $E_{\rm IC, bk}$ 
    (see equations~\ref{eqn:E_ebk} and \ref{eqn:E_synbk}).  The shaded grey band, spanning from $0.1\,E_{\rm syn, bk}$ to $10\, E_{\rm syn, bk}$, is used to evaluate $F_{\rm syn}$ in Section~\ref{subsec:stacking}. The spectral energy distribution plots zoomed in at energies between 
    10~GeV and 100~TeV are presented in Appendix $\ref{subsec:zoomedseds}$ (Figure~\ref{fig:sedzoom}). 
    } 
\end{figure*}

Figure~\ref{fig:sed} shows the spectral energy distribution of our target sources ranging from X-rays to multi-TeV gamma-rays. LS~5039 is currently the only source in our list that has been detected at TeV energies \citep{Mariaud:2015pha}. Our limits below 10~TeV are consistent with the observation of this source by the IACTs. For CYG~X-1 and CYG~X-3, the upper limits from MAGIC \citep{2017MNRAS.471.3657Z} and VERITAS \citep{Archambault:2013yva} are more constraining at 1 TeV but approach the HAWC upper limits as the energy goes up. Finally for SS~433, our limits are slightly less constraining in the first quasi-differential bin but become comparable to the combined MAGIC-H.E.S.S. data \citep{Ahnen:2017tsc} at higher energies. This could be due to a potential contribution from the SS~433 west lobe \citep{Abeysekara:2018qtj}, which is not included in the 3HWC Catalog \citep{2020arXiv200708582A}.

In Figure~\ref{fig:sed}, containment bands are displayed to indicate the HAWC sensitivity at each location. A point source model is fitted in the empty regions of the sky along the same declination band as the target HMMQ to calculate the expected upper limits containing $68\%$ and $95\%$ in yellow and green, respectively. For the calculation of the sensitivities, regions with VHE gamma-ray sources such as the Galactic plane have been excluded. Indeed our upper credible intervals, in red, are at most about 2 $\sigma$ above the expected HAWC limit if there were no emission (dashed black line). Hence, we do not have a clear detection of the HMMQs.

For the four sources discussed in this work, HAWC provides the most stringent upper limits above 10~TeV.

\subsection{Stacking Analysis}
In neither stacking analysis does the combination of the four sources result in a significant detection.  However, the stacked flux limits allow us to set limits on parameters of the scenarios. 

In scenario I, we find the best-fit flux norm $K\,\sum_i \,C_i = (2.4\pm1.1_{\rm stat} \pm 0.4_{\rm sys})\times 10^{-15}\,\rm {TeV^{-1}\,cm^{-2}\,s^{-1}}$ with $\rm {TS}=4.4$ and the best-fit spectral index $\hat{p}=2.2$. It corresponds to a 95\% Confidence Interval (C.I.) limit on the stacked flux $\Phi_\gamma(E_{\rm piv}) = 4.8\times 10^{-15}\,\rm {TeV^{-1}\,cm^{-2}\,s^{-1}}$.  
Through equation~\ref{eqn:phi_gamma1}, we obtain the limit on the jet emission efficiency above 1~TeV,

\begin{equation}\label{eqn:eps_gamma}
    \epsilon_\gamma^{\rm UL} = 5.4\times 10^{-6}.     
\end{equation}
This TeV emission efficiency is 3--5 orders of magnitude lower than the emission efficiency of HMMQs in 0.5--10~keV X-rays, which typically reaches $10^{-3}-10^{-1}$ \citep{1998A&A...338L..71M, 2006A&A...446..591C}. 
We note that $\epsilon_\gamma^{\rm UL}$ is derived using the jet power  in Table~\ref{tab:RHMXB}. The $L_{\rm jet}$ values of the microquasars obtained by different works may differ by a factor of $\sim$2--4 (e.g., \citet{2006AA...451..259P} and \citet{Casares:2005ig}). Note that the uncertainty in $L_{\rm jet}$ is not accounted for in the calculation of $\epsilon_\gamma^{\rm UL}$. 

Our TeV $\gamma$-ray emission efficiency constrains the high-energy neutrino emission efficiency $\epsilon_\nu$ of HMMQs. If VHE gamma rays are produced by the decay of neutral pions, the same proton-proton interaction should produce charged pions that decay into high-energy neutrinos with an emission efficiency $\epsilon_\nu\approx 3\epsilon_\gamma/2$ \footnote{Photopion production is not expected to happen in the stellar radiation field of HMMQs. This is because the $\Delta$-resonance occurs at  
$E_{\rm p, thr} \approx (E_{\Delta} / 2\, \epsilon_0)\, m_p c^2 = 47\, (\epsilon_0/3\,\rm eV)^{-1}\,\rm PeV$, which is above the maximum acceleration energy of the binary based on the Hillas criteria, $E_{p,\,\rm max}< e\,B\,d = 9\,(B/20\,{\rm G})\,(d/0.1\,{\rm AU})\,\rm PeV$, where $E_\Delta \approx 0.3\,\rm GeV$ and $\epsilon_0$ is the typical energy of photons from the companion star.}. The $\epsilon_\gamma$ derived in equation~\ref{eqn:eps_gamma} suggests that a mean-orbital $\epsilon_\nu\sim 0.2$ assumed by   \citet{2006PhRvD..73f3012C} is overly optimistic. The emission efficiency also implies that neutrino detection of HMMQs is difficult with the current neutrino detectors \citep{2018arXiv181107979I}.

In scenario II, using the model described in $\S\ref{subsec:stacking}$ and the INTEGRAL and COMPTEL observations of the sources, $F_{\rm obs, bk}$, the stacking analysis yields the best-fit flux norm, $K\,\sum_i \,C_i = (6.0\pm8.8_{\rm stat}\pm 0.7_{\rm sys})\times 10^{-16}\,\rm {TeV^{-1}\,cm^{-2}\,s^{-1}}$ with $\rm {TS}<4$ and the best-fit spectral index $\hat{p}=2.1$.  The 95\% C.I. upper limit on the stacked $\gamma$-ray flux is   $\Phi_\gamma(E_{\rm piv}) = 2.4\times 10^{-15}\,\rm {TeV^{-1}\,cm^{-2}\,s^{-1}}$, which  
corresponds to a lower limit on the magnetic field strength,

\begin{equation}
    B^{\rm LL} = 22 \,\left(\frac{\epsilon_{\rm syn}}{10\,\%}\right)^{1/2} \,\rm G, 
\end{equation}
where $\epsilon_{\rm syn}$ is an unknown factor denoting the ratio of the actual synchrotron emission by the electron population that emits VHE gamma rays to the total observed 10~keV--10~MeV flux, $\epsilon_{\rm syn} \equiv F_{\rm syn} / F_{\rm obs, bk}$. 

To evaluate the dependence of our result on the $\gamma$-ray spectrum and consider that the $\gamma$-ray spectrum may not strictly follow a power-law spectrum, we also perform the analysis with a log-parabolic spectral model,

\begin{equation}\label{eqn:phi_gamma_logparabola}
    \Phi_\gamma = \frac{F_{\rm syn}\,u_0\,f_{\rm KN}}{u_B}\,K_l\,\left(\frac{E}{E_{\rm piv}}\right)^{-\alpha_l-\beta_l\log(E/E_{\rm piv})}.
\end{equation}
We fix $E_{\rm piv}$ at 7~TeV and scan the indices, $\alpha_l$ between 2.0 and 5.0 with 0.5 interval, and $\beta_l$ between 0.1 and 2.1 with 0.4 interval to find the best-fit flux norm $K_l$. The fit to the data however does not significantly improve with an extra parameter. The lower limit on the magnetic field strength is obtained to be $B^{\rm LL}_{l} = 15\,(\epsilon_{\rm syn}/10\%)^{1/2}$~G, which is comparable to the limit from the power-law assumption.

The derived magnetic field strength agrees with the finding of \citet{2015A&A...581A..27D}, where $B\approx 20$~G was obtained by fitting a relativistic hydrodynamics model to the multi-wavelength observation of LS~5039. \citet{2015A&A...581A..27D} concludes that a high $B$ is unavoidable to explain the COMPTEL flux level of LS~5039. Our result extends the conclusion to all HMMQs accessible to HAWC, and suggests that the large gap between the energy flux in 10~keV--10~MeV and in VHE gamma rays could be a universal feature of HMMQs. Such a high magnetic field challenges the existing models of $\gamma$-ray binaries \citep{2008A&A...489L..21B}, and suggests that the synchrotron component is a small fraction, $\epsilon_{\rm syn} \lesssim 10\%$ of the observed flux between 10~keV and 10~MeV.    
A few caveats should however be noted when interpreting this result as discussed in $\S\ref{sec:discussion}$.

\section{Conclusions and Discussion}\label{sec:discussion}

The highest-energy behaviors of the ``mini'' quasars in our Galaxy are poorly understood, despite the observational and theoretical indications that they provide plausible particle acceleration sites  \citep{Marcote:2015abc,Mariaud:2015pha, Abeysekara:2018qtj}. A lot of microquasars are located close to bright and extended TeV sources, making their observations challenging. By fitting and removing background sources from the regions of interest observed by the HAWC observatory, we provide the most stringent limits on the $\gamma$-ray emissions from LS~5039, CYG~X-1, CYG~X-3 and SS~433 above 10~TeV. By stacking the chance of excess emission from all HMMQs accessible to HAWC, we derive an upper limit of the $\gamma$-ray emission efficiency of HMMQs above 1~TeV, which also constrains the high-energy neutrino emission efficiency of these sources. A second stacking search, applying a standard $\gamma$-ray binary model, further allows us to tightly constrain the contribution of synchrotron emission by relativistic electrons between 10~keV and 10~MeV.

The emission mechanism of hard X-rays / MeV gamma rays from HMMQs has been under debate since the detection of HMMQs by INTEGRAL and COMPTEL (e.g., \citealt{2006A&A...446..591C, Hoffmann:2008yi}). The data can be explained both by thermal Comptonization models where thermal electrons on the accretion disk are Compton scattered \citep{2006A&A...446..591C}, and by non-thermal models where relativistic electrons in the jets produce synchrotron radiation. Our findings challenge the dominance of the latter scenario, and imply the existence of additional emission components or emission zones, especially in the medium $\gamma$-ray band where thermal models become difficult. 

Our model does not account for the $\gamma\gamma$ absorption by the stellar photon field. 
As shown in Appendix~\ref{subsec:tau_pair}, the optical depth of pair production at 1~TeV in the four sources could reach $\sim$10 for head-on interactions but is negligible for tail-on interactions (when the VHE photons are emitted away from the star). As the compact object revolves around the companion, $\gamma\gamma$ absorption could reduce the time-averaged intrinsic $\gamma$-ray flux by a factor of unity. With the attenuation effect, the observed flux would be a fraction of the intrinsic flux, $F_{\rm IC}^{\rm obs}\sim F_{\rm IC}\,\eta_{\gamma\gamma}$, and the magnetic field limit would decrease to  $B^{\rm LL}\,\eta_{\gamma\gamma}^{1/2}$. The fraction $\eta_{\gamma\gamma}$ depends on the poorly-known inclination angle of the binary system. For reference, $\eta_{\gamma\gamma}\sim 0.1-0.4$ is evaluated for the VHE flux of LS~5039 \citep{2006A&A...451....9D}.
Electromagnetic cascades initiated in the pair production process could lead to secondary electrons that emit additional X-ray synchrotron emission, which would further deepen the tension found by our analysis. Our model assumes non- or mildly relativistic outflows like the jets of SS~433. If jets or pulsar winds have a Lorentz factor of a few, $\Gamma > 1$ \citep{2015A&A...581A..27D}, the synchrotron and inverse Compton fluxes would be boosted by the same factor, though the peak energy could be up to $\Gamma$ times higher than the values we estimate. For this reason, we have used a wide energy window to evaluate the fluxes in our second stacking analysis. A more realistic model considering specific outflow configurations is however beyond the scope of this paper. 

Gamma-ray binaries are known to exhibit periodic modulation in flux consistent with their intrinsic properties such as orbital periods. Although the mechanism itself is not fully understood, all of the identified gamma-ray binaries thus far have had their orbital modulations observed \citep{2013A&ARv..21...64D}. For each of the four HMMQs being analysed, we looked for signs of periodic modulations in flux using the HAWC data subdivided into one-transit (daily) maps. Upon adopting Lomb-Scargle periodograms \citep{1982ApJ...263..835S} 
for the time-dependent analysis, no periodic signals were identified. Future VHE gamma-ray observatories such as the Large High Altitude Air Shower Observatory (LHAASO, \citealt{2019arXiv190502773B}) the Southern Wide-Field Gamma-Ray Observatory (SWGO, \citealt{2019BAAS...51g.109H}) and Cherenkov Telescope Array (CTA; \citealt{2019scta.book.....C}) will provide better sensitivities to study the phase-dependent TeV gamma-ray emission from microquasars.



\bigskip
We acknowledge the support from: the US National Science Foundation (NSF); the US Department of Energy Office of High-Energy Physics; the Laboratory Directed Research and Development
(LDRD) program of Los Alamos National Laboratory; Consejo Nacional de Ciencia y Tecnolog{\'i}a
(CONACyT), M{\'e}xico, grants 271051, 232656, 260378, 179588, 254964, 258865, 243290, 132197, A1-S-46288, A1-S-22784, c{\'a}tedras 873, 1563, 341, 323, Red HAWC, M{\'e}xico; DGAPA-UNAM grants
IG101320, IN111315, IN111716-3, IN111419, IA102019, IN112218; VIEP-BUAP; PIFI 2012, 2013,
PROFOCIE 2014, 2015; the University of Wisconsin Alumni Research Foundation; the Institute of
Geophysics, Planetary Physics, and Signatures at Los Alamos National Laboratory; Polish Science
Centre grant, DEC-2017/27/B/ST9/02272; Coordinaci{\'o}n de la Investigaci{\'o}n Cient{\'i}fica de la Universidad Michoacana; Royal Society - Newton Advanced Fellowship 180385; Generalitat Valenciana,
grant CIDEGENT/2018/034; Chulalongkorn University’s CUniverse (CUAASC) grant; Instituto de F{\'i}sica Corpuscular, Universitat de Val{\`e}ncia grant E-46980; National Research Foundation of Korea grant 2018R1A6A1A06024977. Thanks to
Scott Delay, Luciano D{\'i}az and Eduardo Murrieta for technical support.

\bibliography{references}

\clearpage

\begin{appendix}

\section{Model of $\gamma$-ray Emission} \label{subsec:model}

Relativistic electrons in the outflow of the compact object lose energy due to both synchrotron and inverse Compton radiation. 
The target photon field for the inverse Compton process is dominated by the thermal radiation of the companion star. The energy density $u_0$ of the photon field from a star with temperature $T_*$, radius $R_*$, and distance $d_*$ from the compact object can be written as \citep{2013A&ARv..21...64D}:

\begin{eqnarray}\label{eqn:u_0}
    u_0 = \frac{\sigma_{\rm SB}\,T_*^4}{c} \frac{R_*^2}{d_*^2} = 260\,\left(\frac{T_*}{4\times10^4\,\rm K}\right)^{4}\,\left(\frac{R_*}{10\,R_\odot}\right)^2\,\left(\frac{d_*}{0.2\,\rm AU}\right)^{-2}\,\rm erg\,cm^{-3}
\end{eqnarray}
where $\sigma_{\rm SB}$ is the Stefan-Boltzmann constant. As the thermal radiation peaks at 

\begin{equation}
    \epsilon_{0,\rm pk}= 2.8\,k_B\,T_* =  9.6\,\left(\frac{T_*}{4\times10^4\,\rm K}\right)\,\rm{eV},
\end{equation}
the inverse Compton process of electrons above $E_{e,\rm KN}\approx (m_e c^2)^2/(4\epsilon_0)= 6.5\,(k_BT_*/10{\,\rm eV})^{-1}\,\rm GeV$ is in the Klein-Nishina regime. The factor 
$f_{\rm KN}(\gamma_e)$ in equation~\ref{eqn:ratio} accounts for the Klein-Nishina suppression, such that the fluxes of the inverse Compton and synchrotron radiation at the break energy roughly scale as 

\begin{equation}\label{eqn:bk}
    \frac{F_X}{F_\gamma} \approx  \frac{\dot{\gamma}_{\rm syn}}{\dot{\gamma}_{\rm IC}} \approx \frac{u_B}{u_0\,f_{\rm KN}}.
\end{equation} 

For electrons at energy $\gamma_e m_e c^2$ that inverse-Compton scatter a radiation field with differential energy density $du/d\epsilon_0$, the factor reads \citep{2005MNRAS.363..954M}, 

\begin{equation}\label{eqn:FKN}
    f_{\rm KN} = \frac{1}{u_0}\int\,d\epsilon_0\,F_{\rm KN}(b) \frac{du}{d\epsilon_0}
\end{equation}
where $b \equiv {4\gamma_e \epsilon_0}/{(m_ec^2)}$, $u_0 = \int d\epsilon_0\,du/d\epsilon_0$, and 

\begin{equation}
    F_{\rm KN}(b) = \frac{9}{b^3}\left[\left(\frac{b}{2}+6+\frac{6}{b}\right)\ln(1+b)-\left(\frac{11}{12}b^3+6b^2 + 9b +4\right) \frac{1}{(1+b)^2} - 2 + 2\,{\rm Li}_2(-b)\right]
\end{equation}
where $\rm {Li}_2$ is the dilogarithm function. 
$F_{\rm KN}(b) \approx  {9}/{(2\,b^2)}\left(\log b- {11}/{6}\right)$ for $b\gg 1$. 
For a thermal distribution, 

\begin{equation}
    f_{\rm KN}\sim F_{\rm KN} (b(\epsilon_{0,\rm pk})) \approx 1.2 \times 10^{-3}\left(\frac{E_e}{1\,\rm TeV}\right)^{-2}\left(\frac{T_*}{4\times 10^4\,\rm K}\right)^{-2}
\end{equation}


The dominant energy loss channel changes from the inverse Compton emission at low  energy to the synchrotron emission at high energy, with $\dot{\gamma}_{\rm syn} = \dot{\gamma}_{\rm IC}$ happening at 

\begin{equation}\label{eqn:E_ebk}
    E_{e,\rm bk} \approx 2.8\,\left(\frac{B}{1\,\rm G}\right)^{-1}\left(\frac{T_*}{4\times10^4\,\rm K}\right)\left(\frac{R_*}{10\,R_\odot}\right)\left(\frac{d_*}{0.2\,\rm AU}\right)^{-1}\, \rm TeV. 
\end{equation}
This electron energy corresponds to a break in the synchrotron spectrum at 

\begin{equation}\label{eqn:E_synbk}
    E_{\rm syn, bk}=0.4\,\,\left(\frac{B}{1\,\rm G}\right)^{-1}\left(\frac{T_*}{4\times10^4\,\rm K}\right)^2\left(\frac{R_*}{10\,R_\odot}\right)^2\left(\frac{d_*}{0.2\,\rm AU}\right)^{-2} \,\rm MeV
\end{equation}
and a break in the inverse Compton spectrum at 

\begin{equation}
    E_{\rm IC, bk}\approx E_{e,\rm bk}. 
\end{equation}

In our analysis we solve equation~\ref{eqn:bk} and \ref{eqn:FKN} numerically. The derived source properties, including $u_0$, $F_{\rm KN}$, $E_{\rm syn, bk}$, and $E_{e,\rm bk}$ are listed in Table~\ref{tab:RHMXB_Derived}. 
 
\begin{deluxetable}{ccccc}
\tablecaption{Derived Source Properties.   \label{tab:RHMXB_Derived}}
\tablewidth{700pt}
\tabletypesize{\scriptsize}
\tablehead{
\colhead{Name} & \colhead{$u_0$} & \colhead{$f_{\rm KN}$ at $E_{e,\rm bk}$\tablenotemark{\scriptsize *}} &  \colhead{$E_{\rm syn, bk}$\tablenotemark{\scriptsize *}} &  \colhead{$E_{\rm e, bk}$\tablenotemark{\scriptsize *}}  \\ 
\colhead{} & \colhead{[$\rm erg\,cm^{-3}$]} & \colhead{} &  
\colhead{[keV]} &   \colhead{[TeV] }
} 
\startdata
    LS 5039 \space & 820 & $4.9\times 10^{-5}$ & 2100 & 6.9  \\
    CYG X-1 \space & 380 & $1.1\times 10^{-4}$ & 1400 & 5.6      \\
    CYG X-3 \space & 2560 & $1.55\times 10^{-5}$ & 4620 & 10.2    \\
    SS 433 \space  & 5.5 & $7.2\times 10^{-3}$ & 6.2 & 0.4  \\
\enddata
\tablenotetext{*}{Derived assuming $B=1$~G.}

\end{deluxetable}


\section{Pair production optical depth} \label{subsec:tau_pair}

\begin{figure}[th]
    \centering
    \includegraphics[width=.49\textwidth]{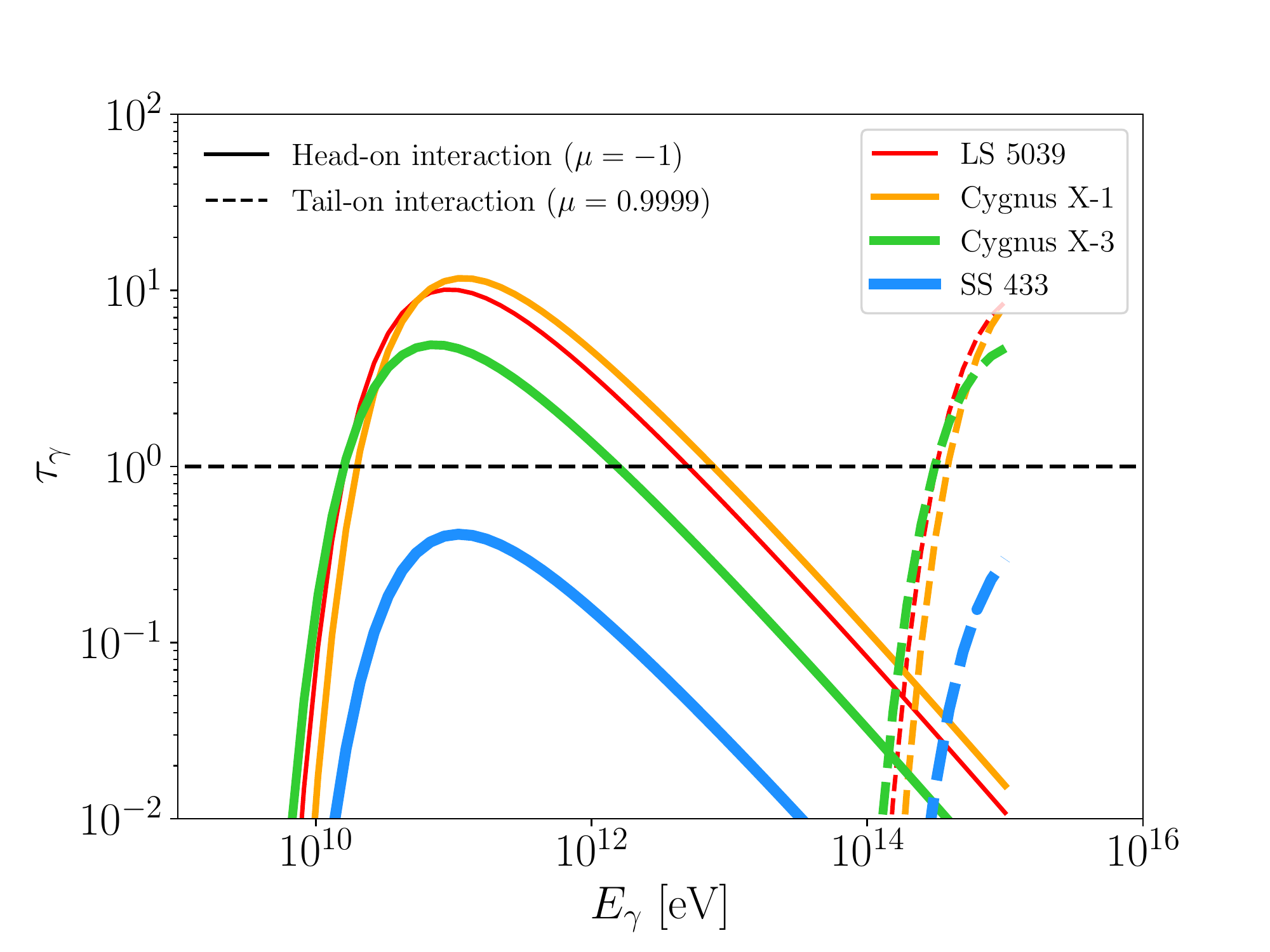}
    \caption{\label{fig:tau_pair} Optical depth of $\gamma\gamma$ pair production with the stellar radiation field of the four sources. The solid and dashed curves correspond to head-on and tail-on interactions, respectively. 
    } 
\end{figure}

A VHE $\gamma$ ray with energy $E_\gamma$ emitted by the jets may interact with stellar photons with energy $\epsilon_0$ through $\gamma\gamma$ pair production. The pair production cross section is 

\begin{equation}
    \sigma_{\gamma\gamma}(s) = \frac{3}{16}\sigma_T (1-\beta_{\rm cm}^2) \left[ (3-\beta_{\rm cm}^4)\ln\left(\frac{1+\beta_{\rm cm}}{1-\beta_{\rm cm}}\right) - 2\beta_{\rm cm}(2-\beta_{\rm cm}^2)\right]
\end{equation}
where $\beta_{\rm cm} = (1-4/s)^{1/2}$ is the centre-of-momentum speed of the incident particles, $s$ is the invariant energy, 

\begin{equation}
    s = \frac{2\,E_\gamma\epsilon_0 (1-\mu)}{m_e^2c^4}. 
\end{equation}
$\mu = \cos(\theta)$ and $\theta$ is the interaction angle between the directions of $E_\gamma$ and $\epsilon_0$.  

The optical depth of the pair production is 

\begin{equation}
    \tau_{\gamma\gamma} (E_\gamma, \mu) \approx \int d\epsilon_0 \frac{dn}{d\epsilon_0} \sigma_{\gamma\gamma}(E_\gamma, \epsilon_0, \mu) d_*.
\end{equation}

Figure~\ref{fig:tau_pair} shows $\tau_{\gamma\gamma}$ for various gamma-ray energies for head-on ($\mu=-1$) and tail-on ($\mu=1-10^{-4}$) interactions. In the head-on case, VHE $\gamma$ rays are heavily attenuated. In the tail-on case, no pair production occurs below $\sim 100$~TeV. As the VHE source revolves around the star, the $\gamma$-ray attenuation effect is between these two extreme cases and depends on the orbital phase. 

\section{Significance Maps} \label{subsec:sigmap}

\begin{figure*}[t!]
    \centering
    \includegraphics[width=.49\textwidth]{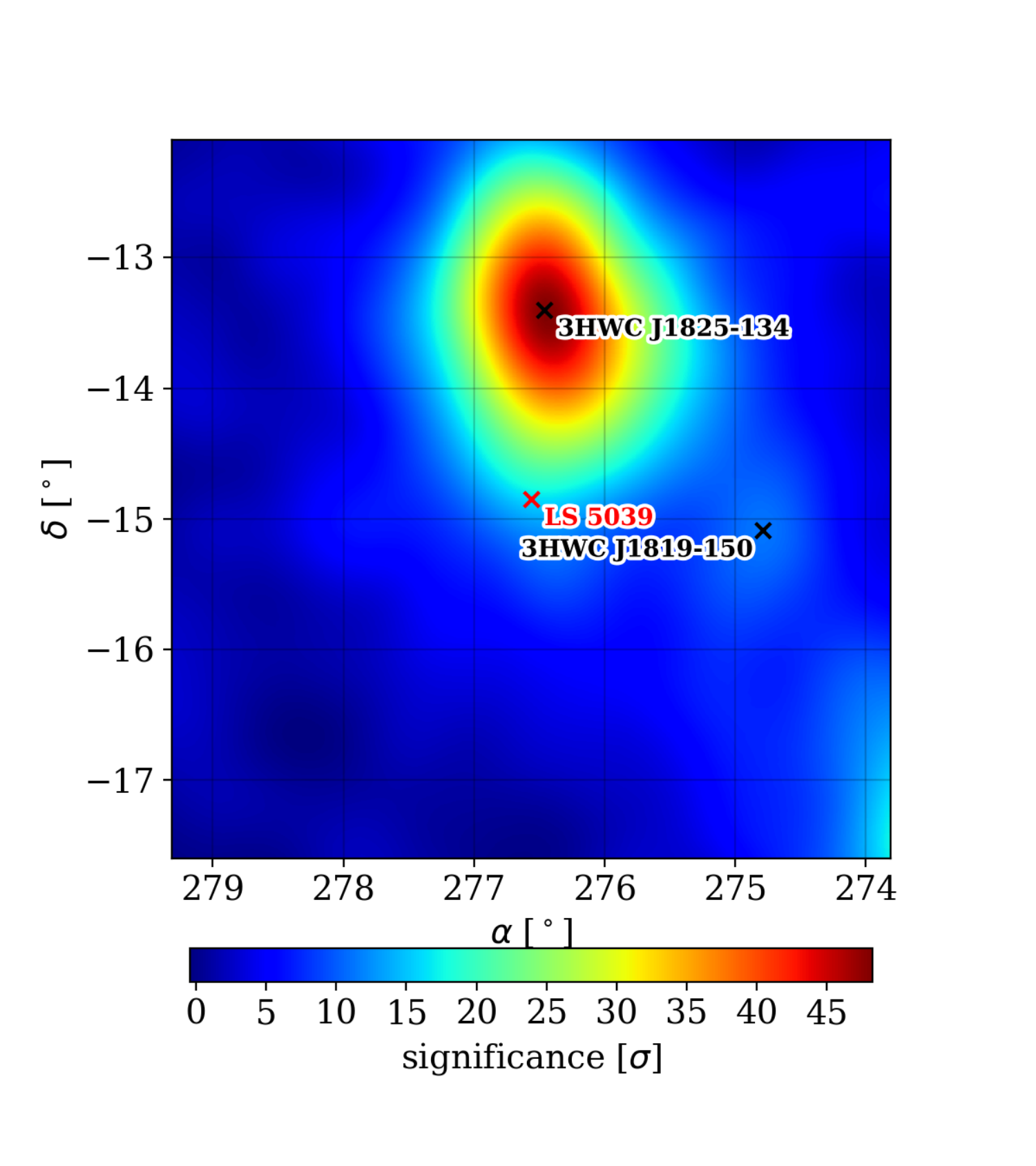}
    \hfill
    \includegraphics[width=.49\textwidth]{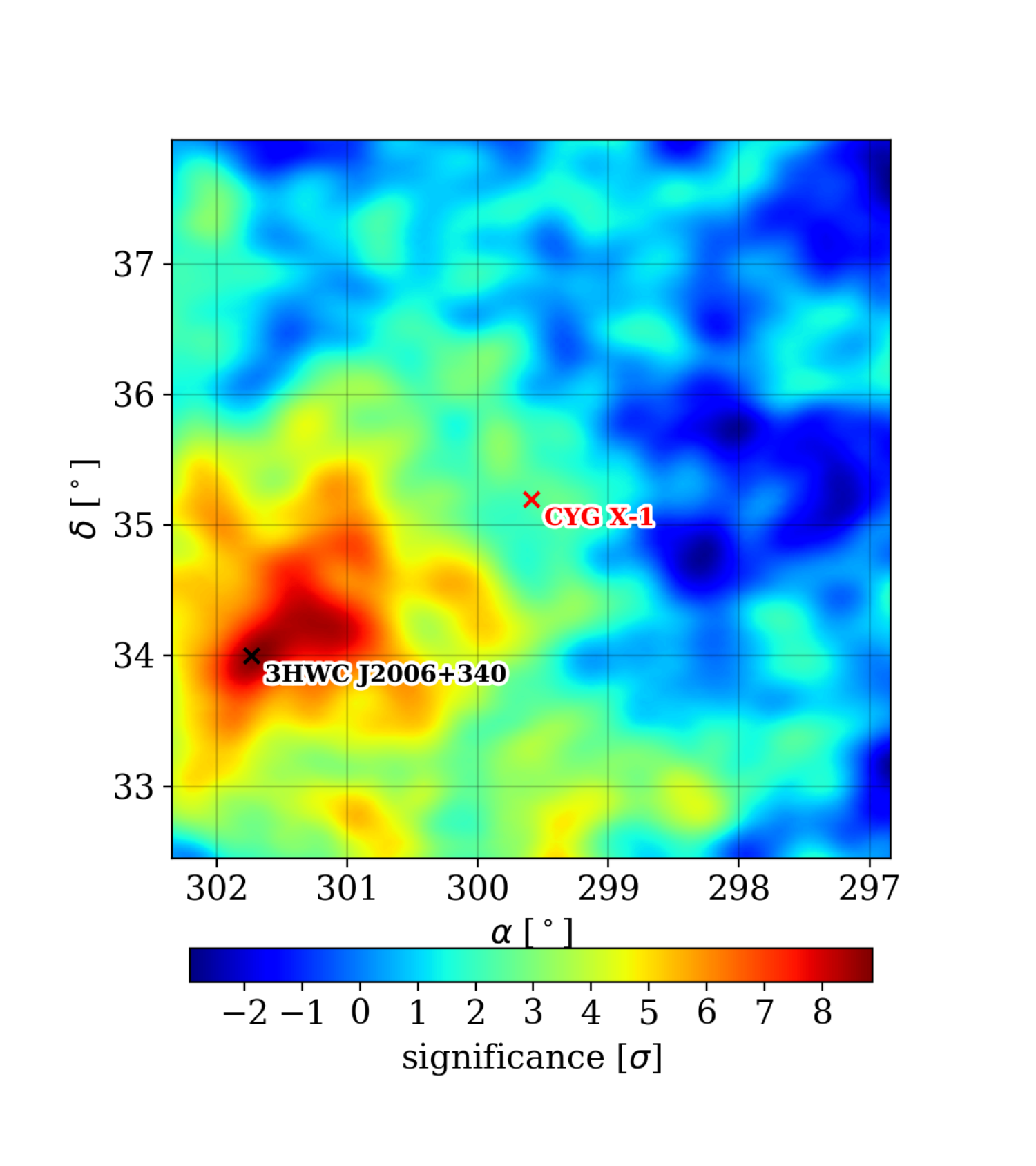}\\
    \includegraphics[width=.49\textwidth]{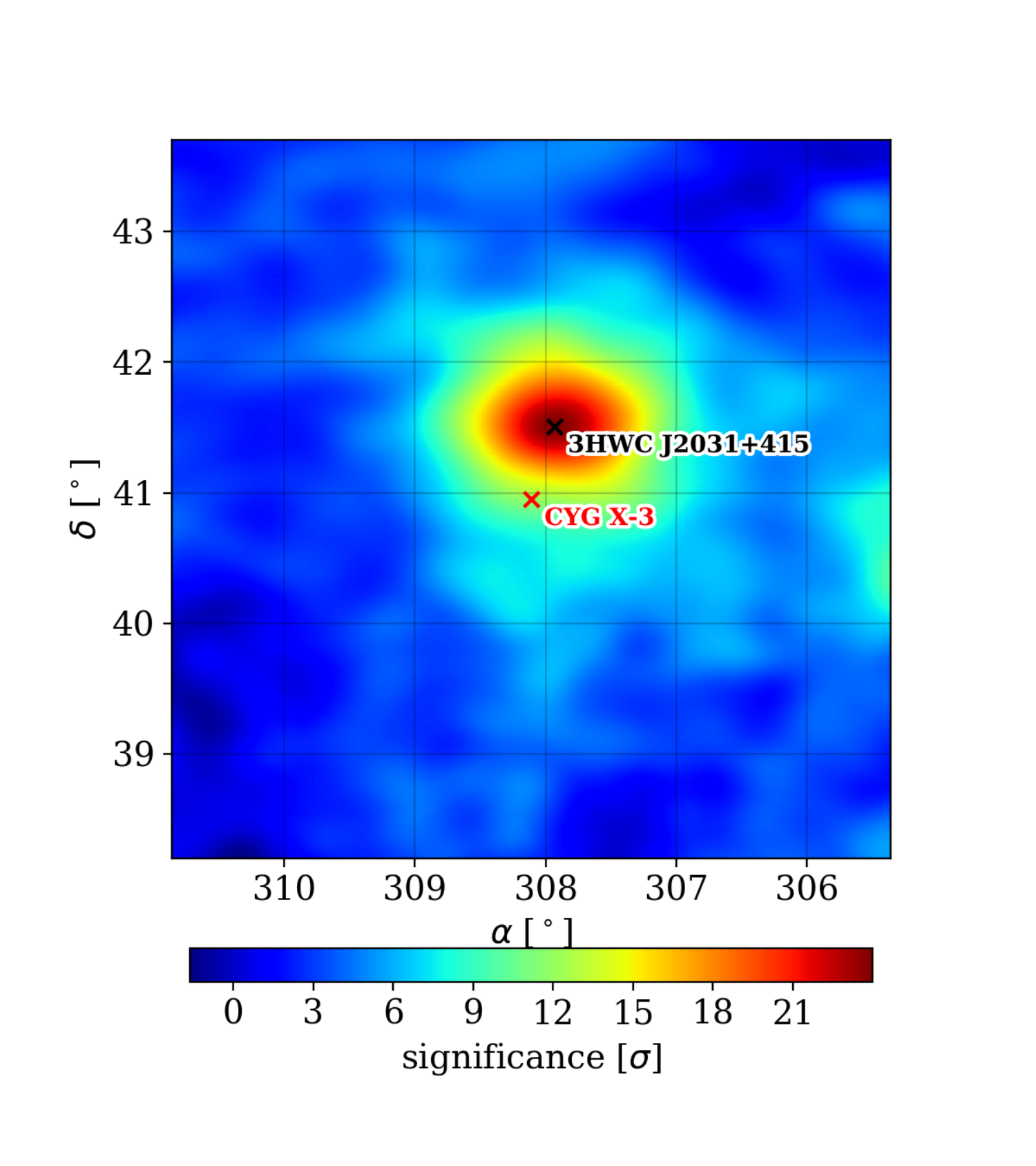}
    \hfill
    \includegraphics[width=.49\textwidth]{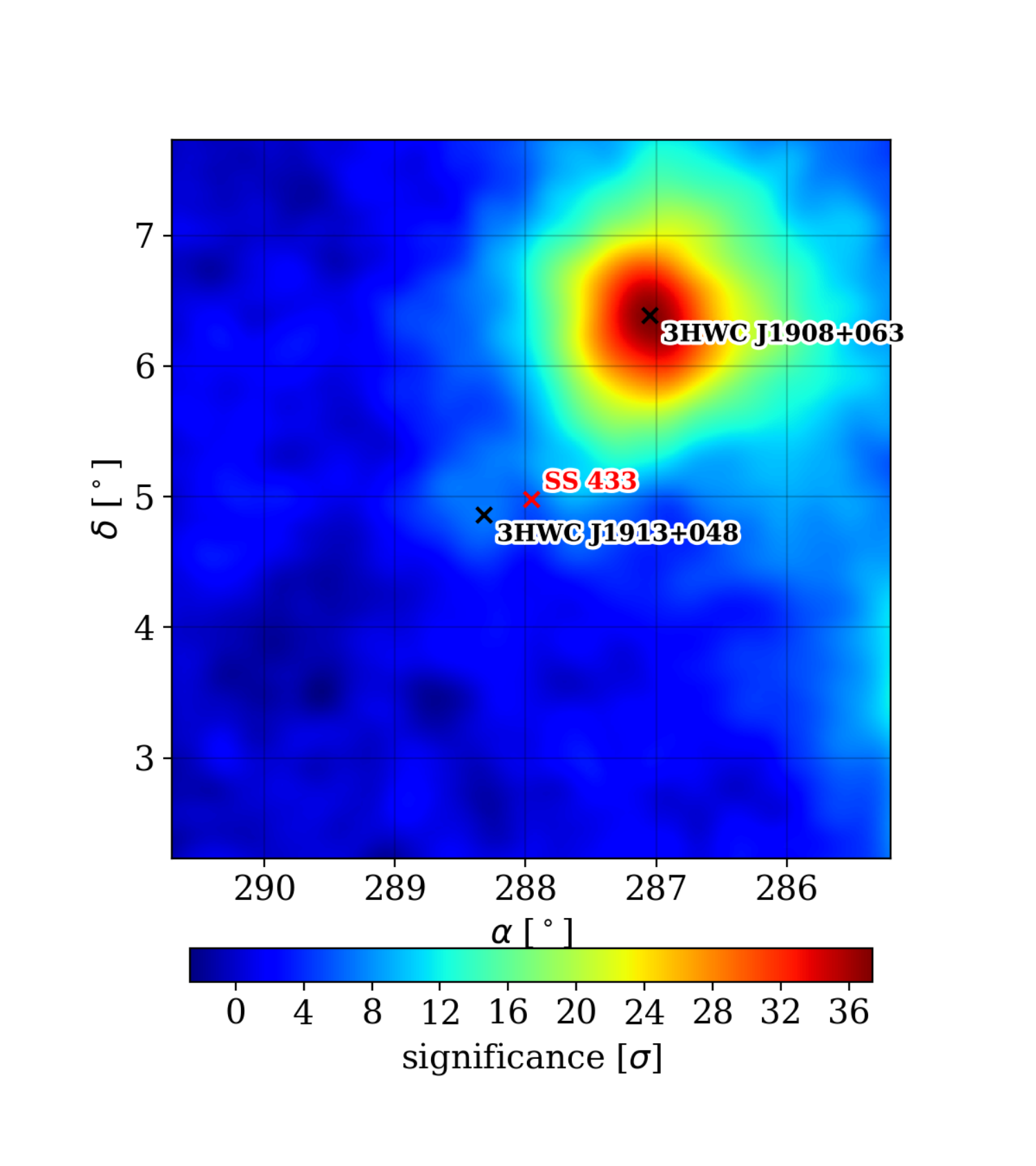}\\
    \caption{\label{fig:sigmap} Significance maps of LS 5039 (top left), CYG X-1 (top right), CYG X-3 (bottom left), and SS 433 (bottom right) produced using 1,523 days of HAWC data.}
\end{figure*}

Figure~\ref{fig:sigmap} presents the significance maps of the regions of the four HMMQs in equatorial coordinates before removing photon counts from nearby 3HWC sources. 

\section{Flux Upper limit with A Single Energy Bin} \label{subsec:zoomedseds}

\begin{figure*}
    \centering
    \includegraphics[width=.49\textwidth]{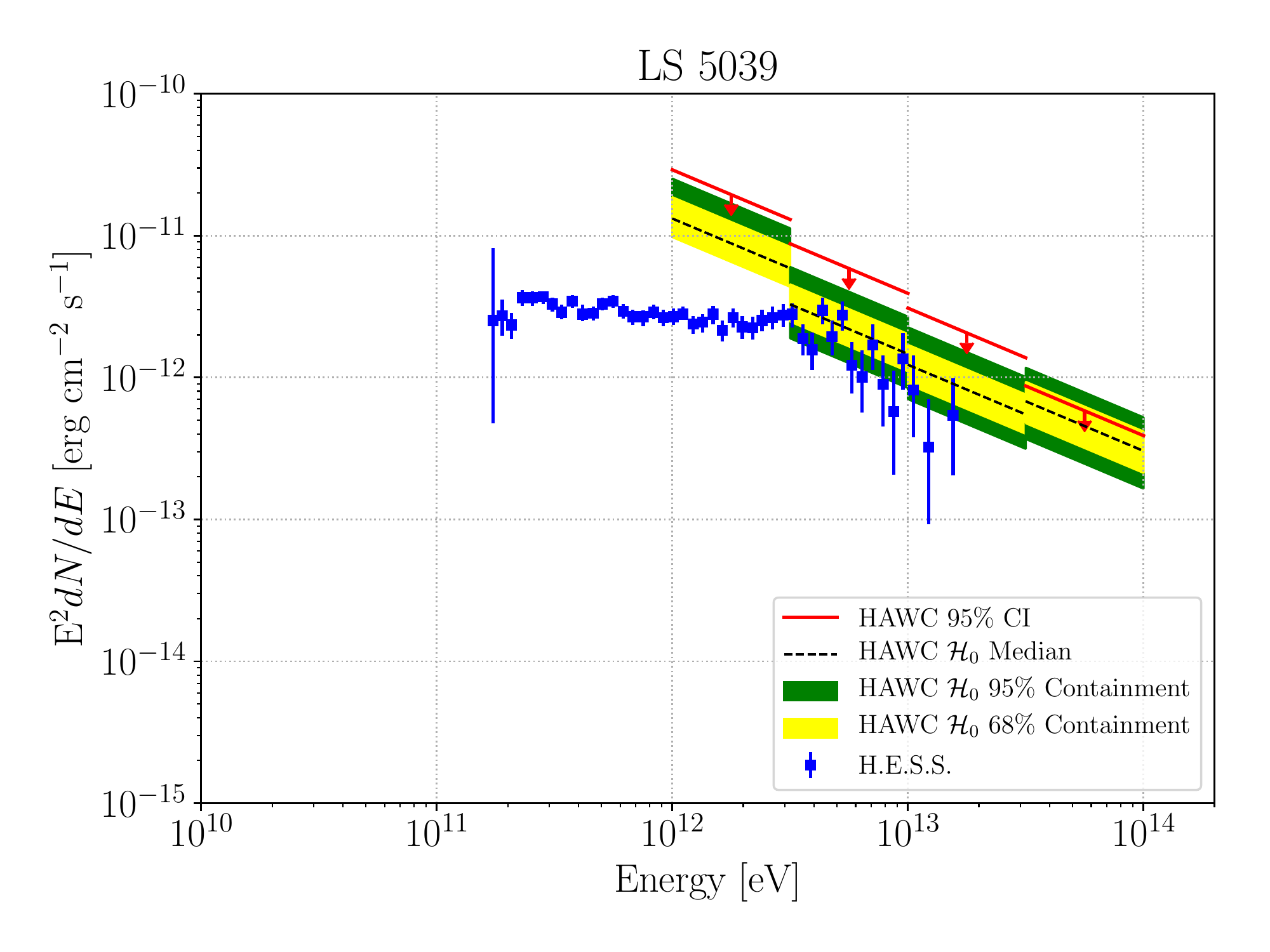}
    \hfill
    \includegraphics[width=.49\textwidth]{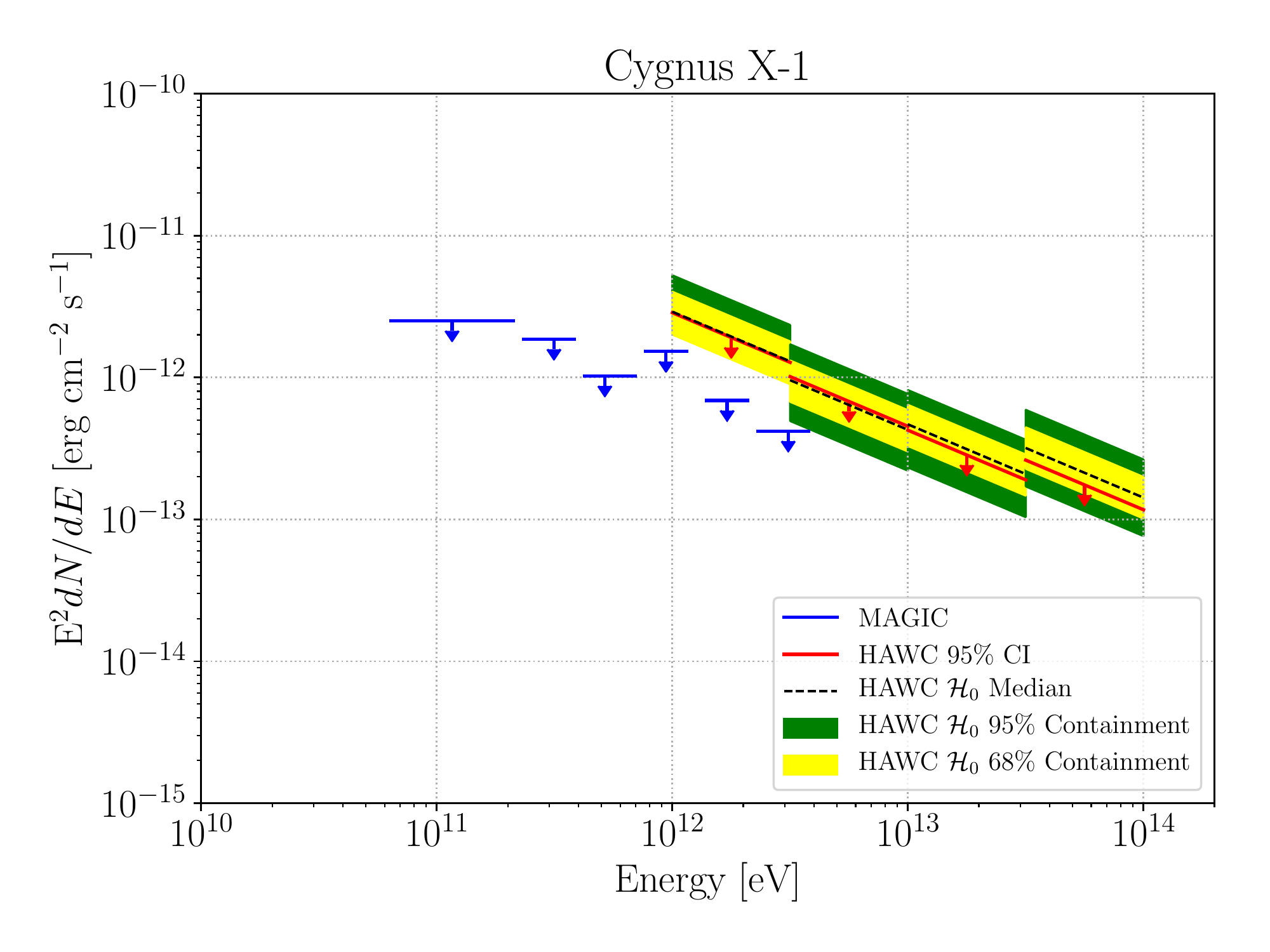}\\
    \includegraphics[width=.49\textwidth]{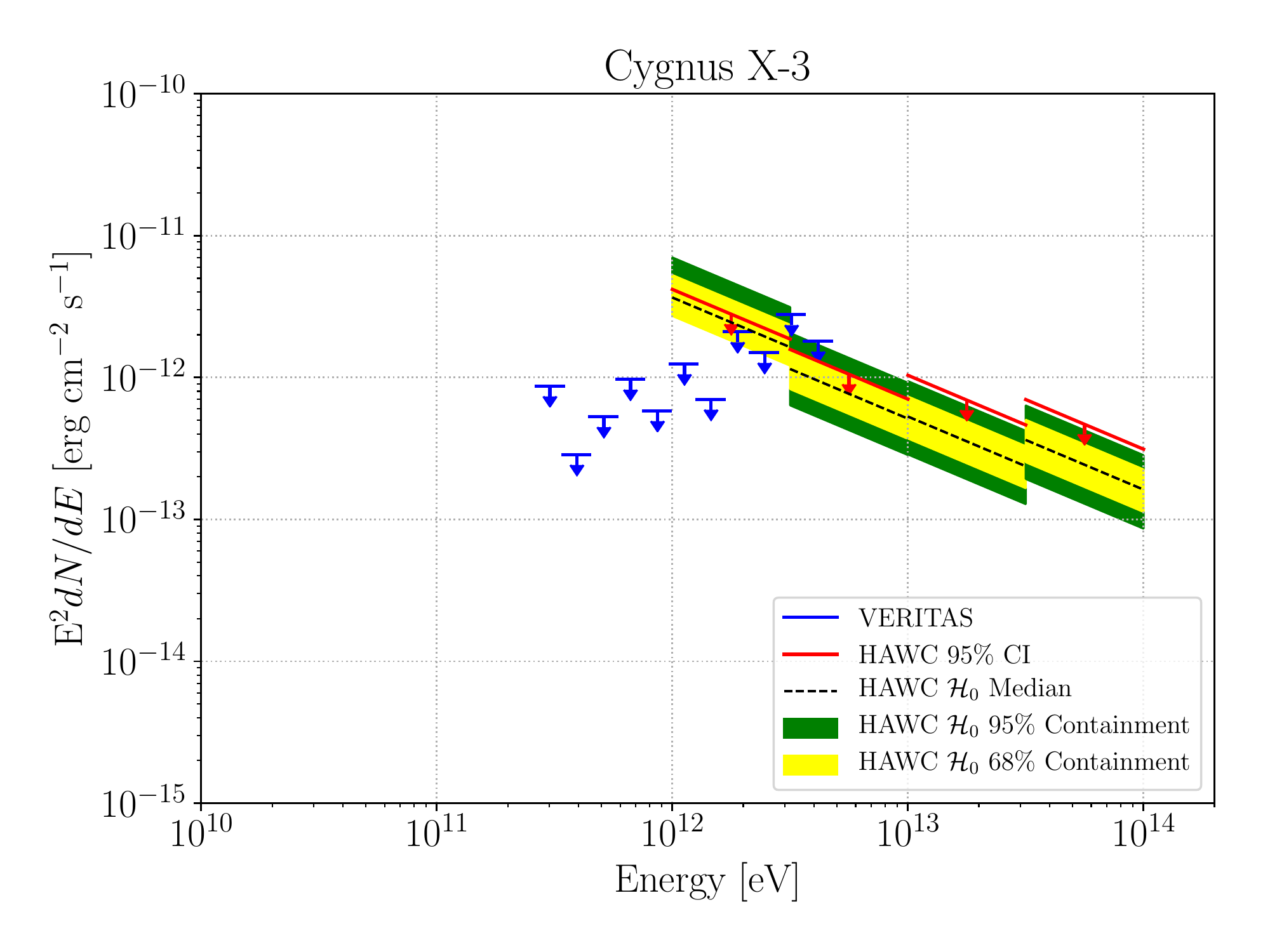}
    \hfill
    \includegraphics[width=.49\textwidth]{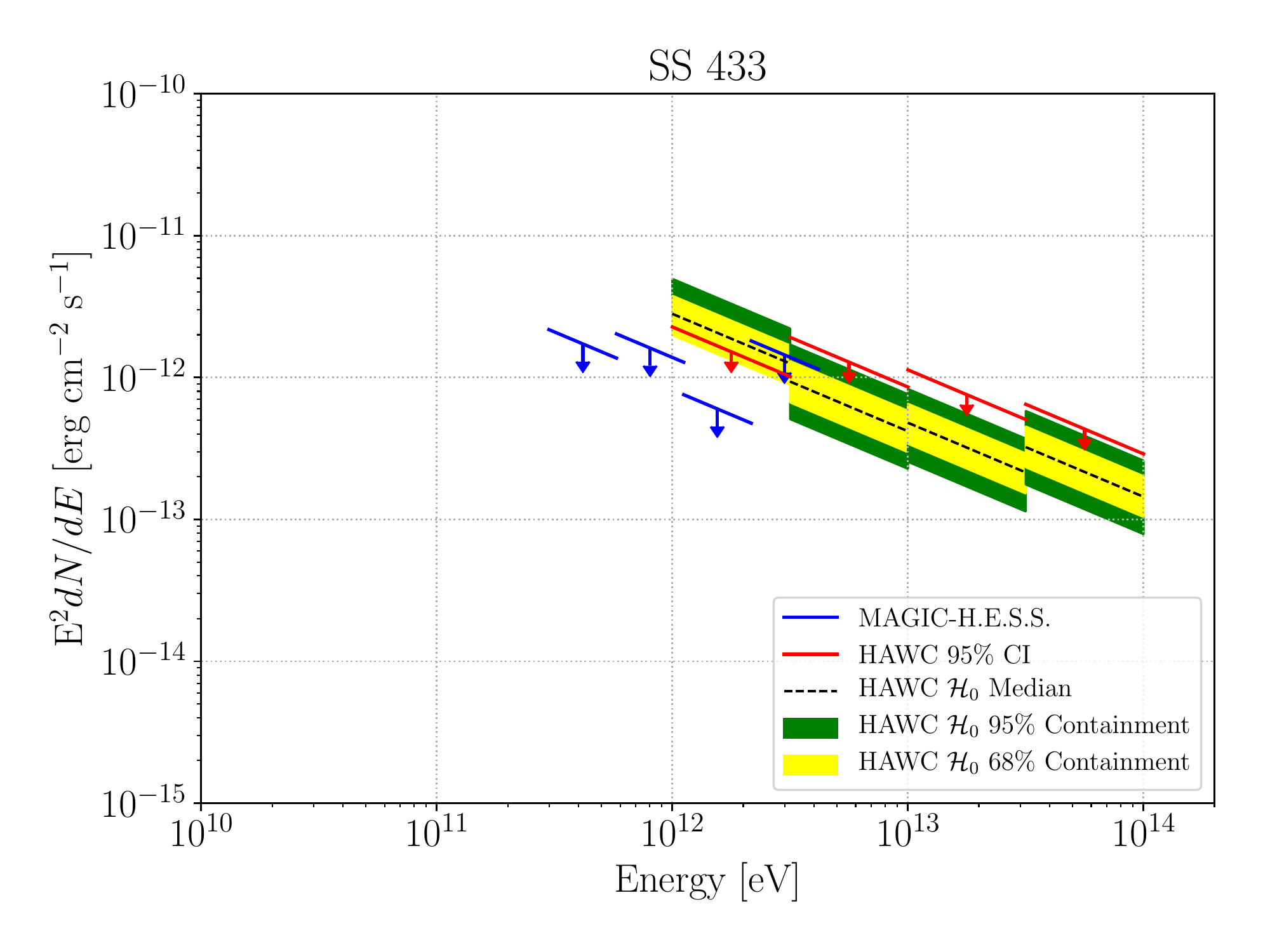}\\
    \caption{\label{fig:sedzoom} Spectral energy distribution of LS 5039 (top left), CYG X-1 (top right), CYG X-3 (bottom left), and SS 433 (bottom right). Features Gamma-ray data from various IACTs in blue in comparison with the upper limits on VHE $\gamma$-rays derived from the HAWC observation. 
    } 
\end{figure*}

Figure~\ref{fig:sedzoom} displays the spectral energy distributions of the four HMMQs between 10~GeV and 200~TeV as a zoom in view of Figure~\ref{fig:sed}.

\begin{figure*}
    \centering
    \includegraphics[width=.49\textwidth]{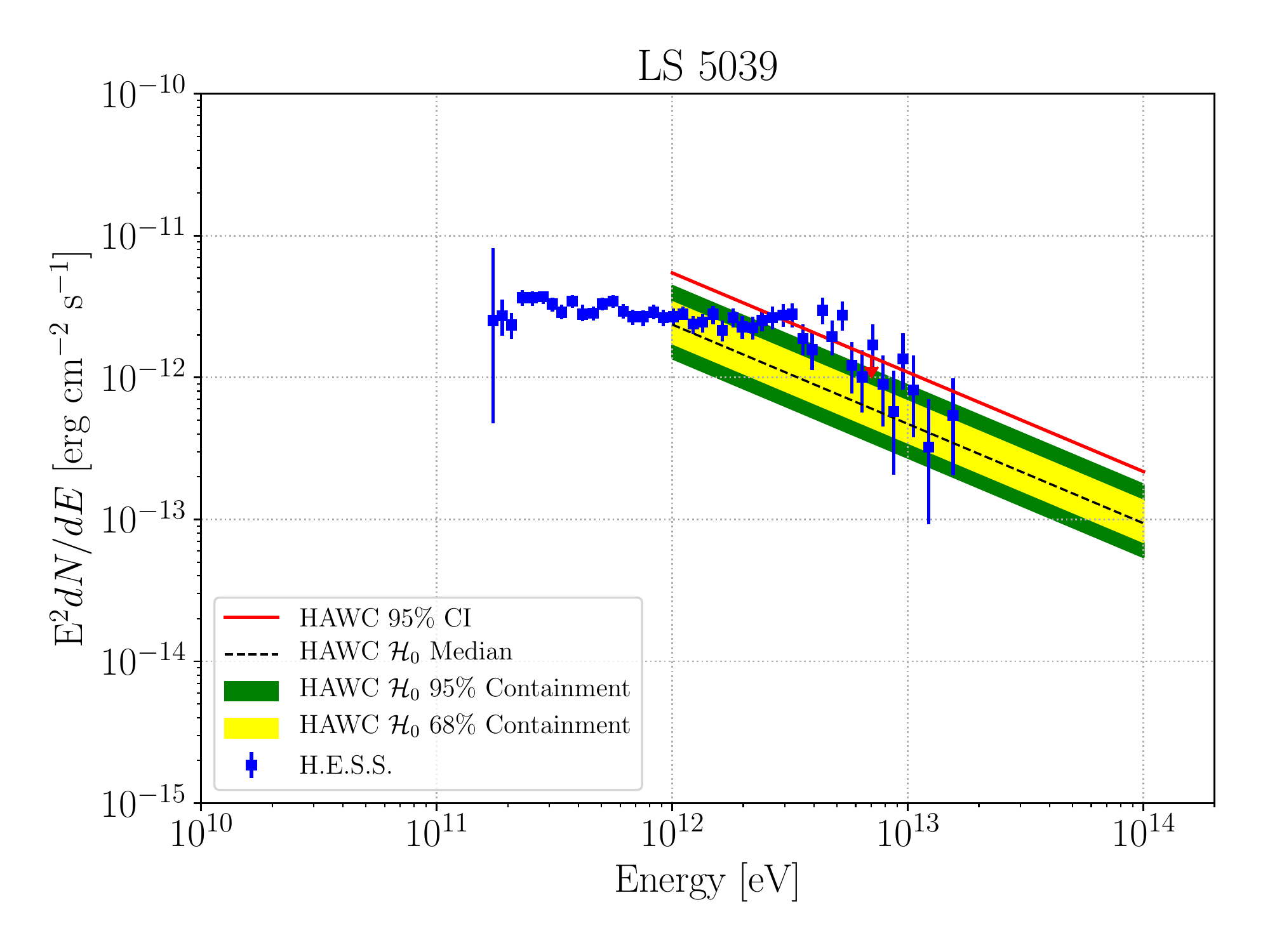}
    \hfill
    \includegraphics[width=.49\textwidth]{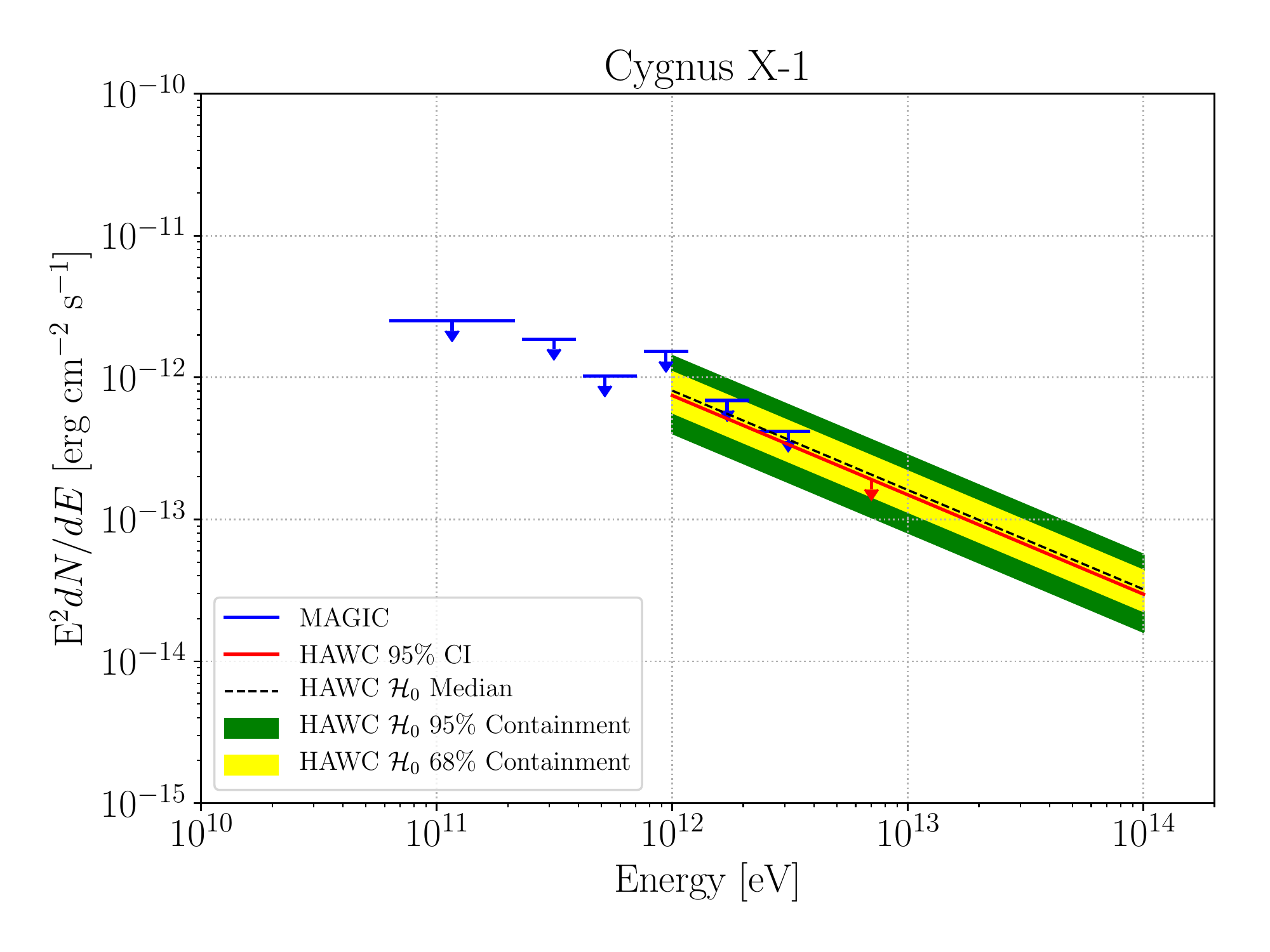}\\
    \includegraphics[width=.49\textwidth]{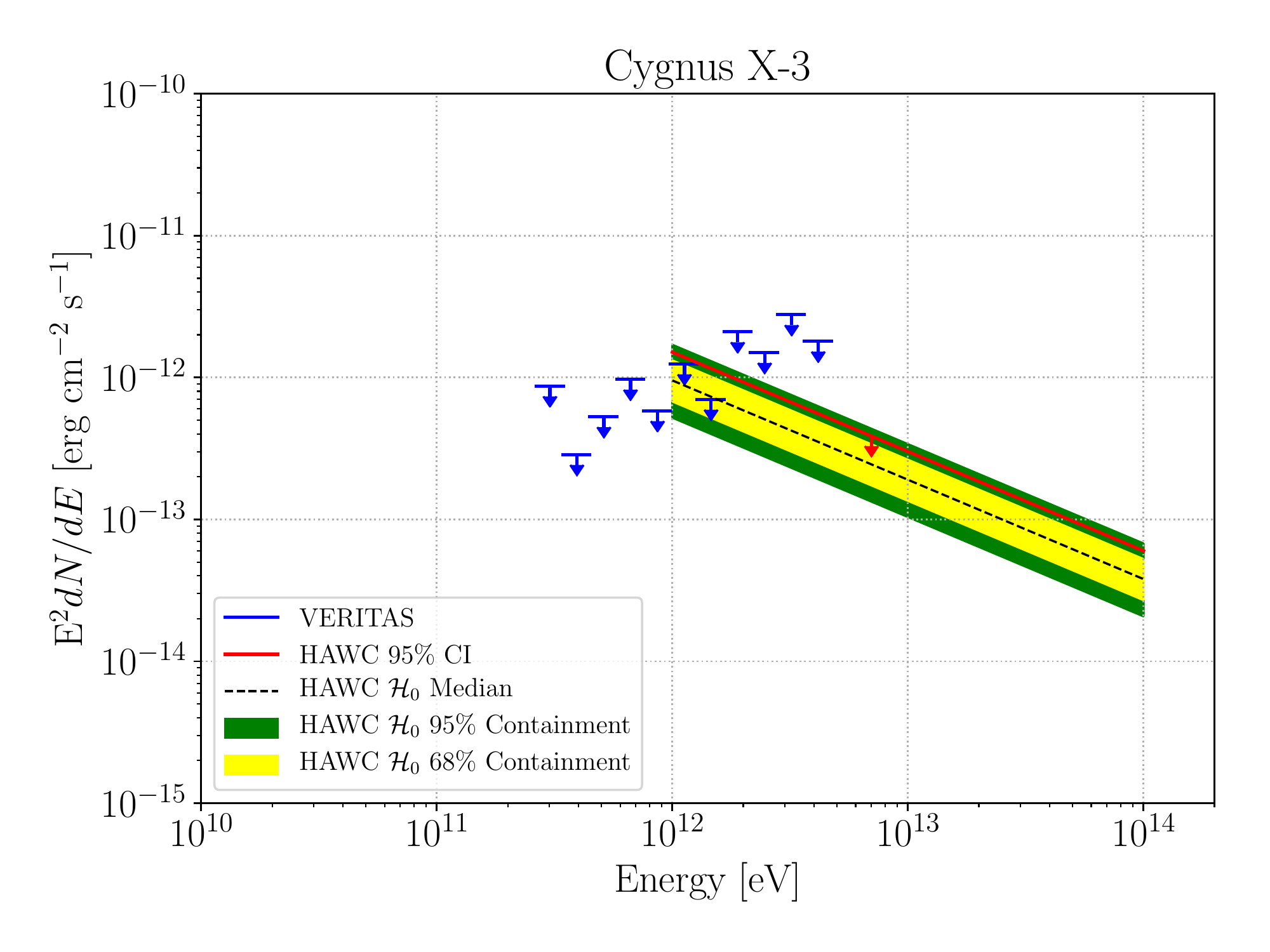}
    \hfill
    \includegraphics[width=.49\textwidth]{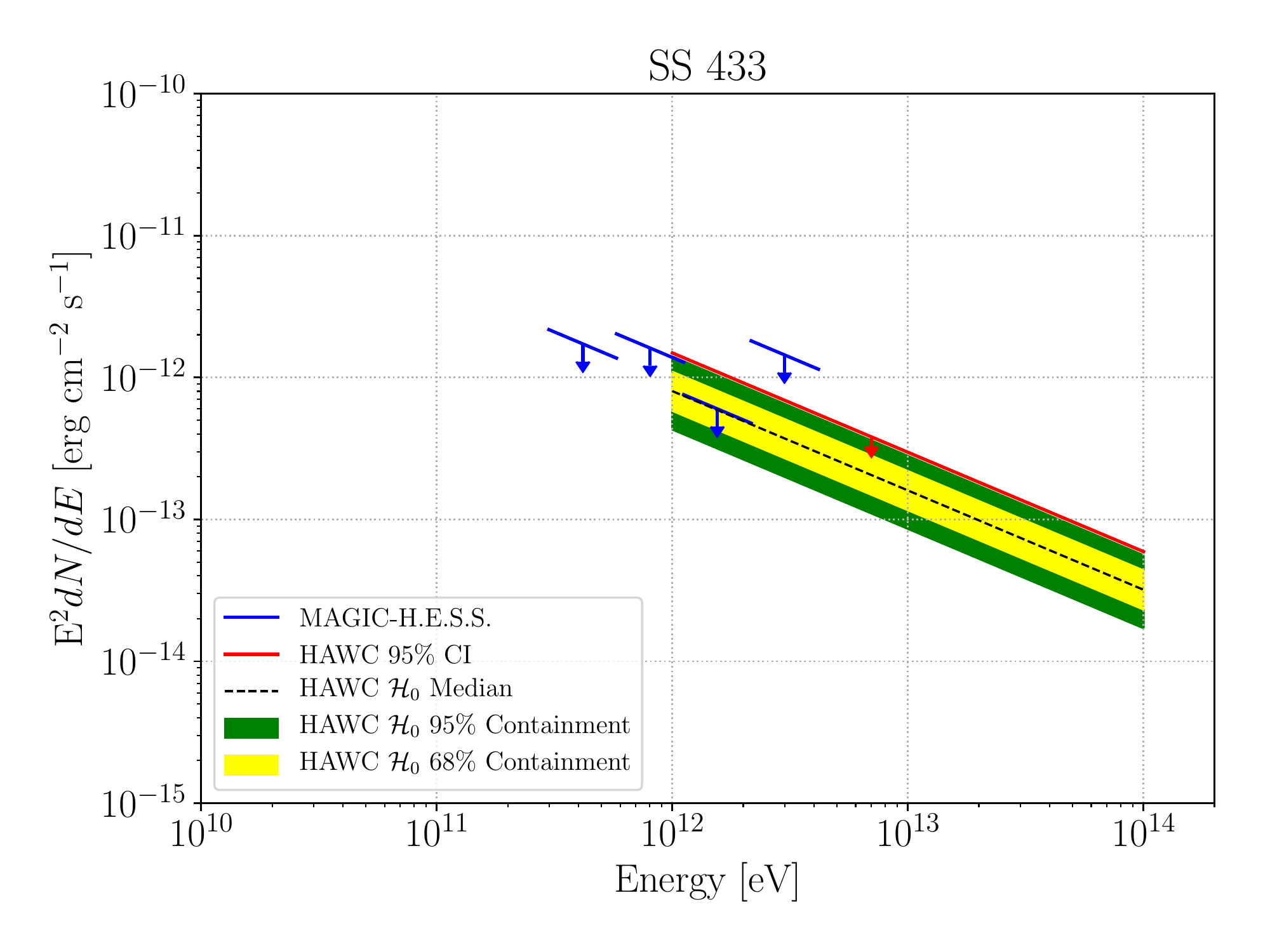}\\
    \caption{\label{fig:sedzoomfullenergy} 
    Same as Figure~\ref{fig:sedzoom} except that the HAWC upper limits and sensitivity bands are computed in the full energy range of HAWC.
    } 
\end{figure*}

In Figure~\ref{fig:sedzoomfullenergy}, we also show the flux upper limits and HAWC sensitivities when using one single energy bin with $E>1$~TeV. These limits are tighter than the differential limits in Figure~\ref{fig:sed}. It is likely due to the larger statistics when combining events from all four energy bins. However, the systematic uncertainty due to the choice of the spectral index is more significant with the single energy bin. We report the associated systematic uncertainties in Table~\ref{tab:sys-quasi}.

\section{Systematic Effects}\label{sec:systematics}
Table~\ref{tab:sys-quasi} shows the systematic effects on the flux normalizations for the HMMQs. We evaluate the impact of two systematic errors. The first is the uncertainty due to the detector response, and the second is due to the choice of the spectral index in our power-law model. The uncertainty due to detector response is at the level of $10-20$\% for most sources and energy bins except the fourth bin of the CYG~X-1 analysis. The statistics for this source above 30~TeV is deficient and the fits are not adequately converged. The uncertainty due to the choice of the spectral index is  $<20$\% with differential bins but rises significantly with a single energy bin.  

\begin{deluxetable}{ccccc}
\tablecaption{Systematic uncertainties due to detector response and the choice of spectral index for quasi-differential bins and for a single full-energy bin, respectively.\label{tab:sys-quasi}}
\tablewidth{700pt}
\tabletypesize{\scriptsize}
\tablehead{
\colhead{Energy Bin} & \multicolumn{4}{c}{Systematics (detector, index)}\\
& \colhead{LS 5039} & \colhead{CYG X-1} & \colhead{CYG X-3} & \colhead{SS 433}
} 
\startdata
        1 & 6\%, 7\% & 12\%, 9\% & 12\%, 18\% & 11\%, 10\%\\
        2 & 15\%, 4\% & 19\%, 7\% & 14\%, 13\% & 13\%, 3\%\\
        3 & 27\%, 11\% & 21\%, 10\% & 12\%, 4\% & 22\%, 4\%\\
        4 & 28\%, 16\% & 86\%, 51\% & 15\%, 5\% & 22\%, 4\%\\
        Full & 19\%, 72\% & 13\%, 53\% & 16\%, 36\% & 10\%, 34\%
\enddata
\end{deluxetable}

\end{appendix}


\end{document}